\documentclass[11pt]{article}
\usepackage[preprint]{acl}
\usepackage{times,latexsym}
\usepackage[T1]{fontenc}
\usepackage[utf8]{inputenc}
\usepackage{microtype,inconsolata}
\usepackage{graphicx,amsmath,amssymb,booktabs,multirow,tabularx}
\usepackage{fancyvrb,fvextra,pifont}
\usepackage[inline]{enumitem}
\newcolumntype{C}{>{\centering\arraybackslash}X}

\title{Auto-AEG: Scalable Data Construction for Open-Vocabulary Audio Event Grounding}
\author{
  Zihan Zhang\thanks{Equal contribution.}\textsuperscript{1}
  \quad
  Xize Cheng\footnotemark[1]\textsuperscript{1}
  \quad
  Wenhao Yan\footnotemark[1]\textsuperscript{2}
  \quad
  Tong Zhang\textsuperscript{1}
  \quad
  Dongjie Fu\textsuperscript{1}
  \quad
  Wenxu Jia\textsuperscript{1}
  \\
  \textbf{Boyun Zhang\textsuperscript{1}
  \quad
  Yongbo He\textsuperscript{1}
  \quad
  Qifei Zhang\textsuperscript{1}
  \quad
  Linjun Li\textsuperscript{3}
  \quad
  Wenshi Chen\textsuperscript{3}
  \quad
  Tao Jin\textsuperscript{1}}\thanks{Corresponding author.}
  \\
  \textsuperscript{1}Zhejiang University
  \quad
  \textsuperscript{2}Tsinghua University
  \quad
  \textsuperscript{3}Meituan
  \\
  \texttt{zihanzhang.ai@zju.edu.cn, jint\_zju@zju.edu.cn}
}
\begin{document}
\maketitle

\begin{abstract}
Large Audio-Language Models (LALMs) reason fluently about sound yet struggle to
localize precisely when events occur, while classical Sound Event
Detection (SED) attains frame-level precision only over a closed label set. At the
intersection of these paradigms lies the task of \textit{Open-Vocabulary Audio Event Grounding}: 
predicting all time intervals of a target sound event described by an arbitrary natural 
language query. Scaling this task requires diverse temporal supervision covering
every occurrence of a queried event, while manually annotating these boundaries
remains costly.

To address this, we introduce \textbf{Auto-AEG}, a scalable pipeline that constructs such
supervision by automatic data construction and model fine-tuning. It pairs programmatically 
synthesized clips, which carry placement-exact ground-truth intervals for supervised cold-start, with
multi-model pseudo-labels on real-world audio that supply the reward signal for
reinforcement learning. Training with this pipeline yields large temporal-localization gains
% (\textbf{+73.9\%} relative mIoU over zero-shot) 
on \textbf{AEGBench},
an independent difficulty-stratified benchmark whose annotations we release, and these gains
generalize to various held-out SED and audio grounding benchmarks. Our results show that
automatically constructed data, coupled with interval-aware reward design,
provides an effective data-side route to expanding the temporal localization
capability of LALMs.

\end{abstract}

\section{Introduction}

Large Audio Language Models
(LALMs)~\citep{deshmukh2023pengi,gong2024ltu,tang2024salmonn,Qwen2-Audio,goel2025audioflamingo3advancing,xu2025qwen3omnitechnicalreport}
align audio encoders with large language models and have become proficient at
describing, classifying, and reasoning about sound. Precisely localizing sound
events in time, however, remains a persistent challenge~\citep{sun2026spotsoundenhancinglargeaudiolanguage,kulkarni2026tempo}.

Conventional sound event detection (SED) provides frame-level predictions but 
typically operates over fixed label sets, whereas large audio-language models (LALMs) 
support open-ended queries yet often struggle with fine-grained temporal localization. 
We study Open-Vocabulary Audio Event Grounding, a task that bridges these paradigms: 
given an audio clip and a natural-language query describing a target event, predict all 
onset-offset intervals in which the event occurs, allowing multiple occurrences and an 
open vocabulary. Interpreting diverse natural-language queries motivates leveraging the 
linguistic and audio-semantic knowledge of pretrained LALMs, rather than relying solely on 
specialized detectors trained with fixed-label supervision.

\begin{figure}[t]
\centering
\includegraphics[width=\columnwidth]{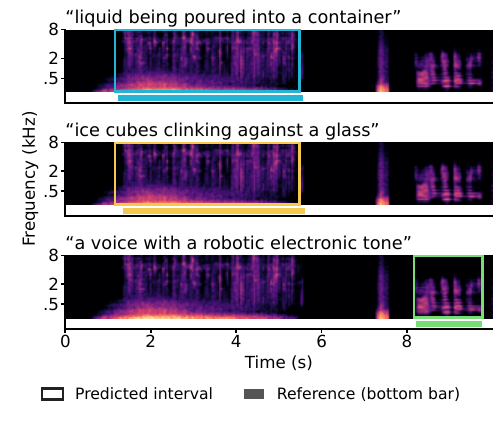}
\caption{Open-vocabulary grounding on one mixture with overlapping sound events. The same mel spectrogram is shown for three natural-language queries. Outlined boxes indicate Auto-AEG GRPO predictions; lower colored bars indicate human-annotated reference intervals.}
\label{fig:qualitative_grounding}
\end{figure}

Figure~\ref{fig:qualitative_grounding} illustrates the task with three descriptive queries on the same recording. Liquid pouring and clinking ice occur concurrently, while a robotic voice appears later; target intervals therefore depend on the query and may overlap across queries. Grounding requires linking linguistic descriptions to sound sources and resolving their temporal extent within mixed audio. By turning free-form descriptions into temporal access points, this task supports precise audio search and navigation through recordings with concurrent sound sources.

Accurate temporal supervision is essential for audio grounding. Existing
resources~\citep{xu2021tag,munakata2025amr,nakada2025castella,primus2025tacos,shao2026listeningwithtime}
provide valuable annotations, but extending them to diverse events and all their
occurrences remains costly. We address this challenge through automatic data
construction for model-transferable post-training without architectural changes.
\begin{figure*}[t]
  \centering
  \includegraphics[width=\linewidth]{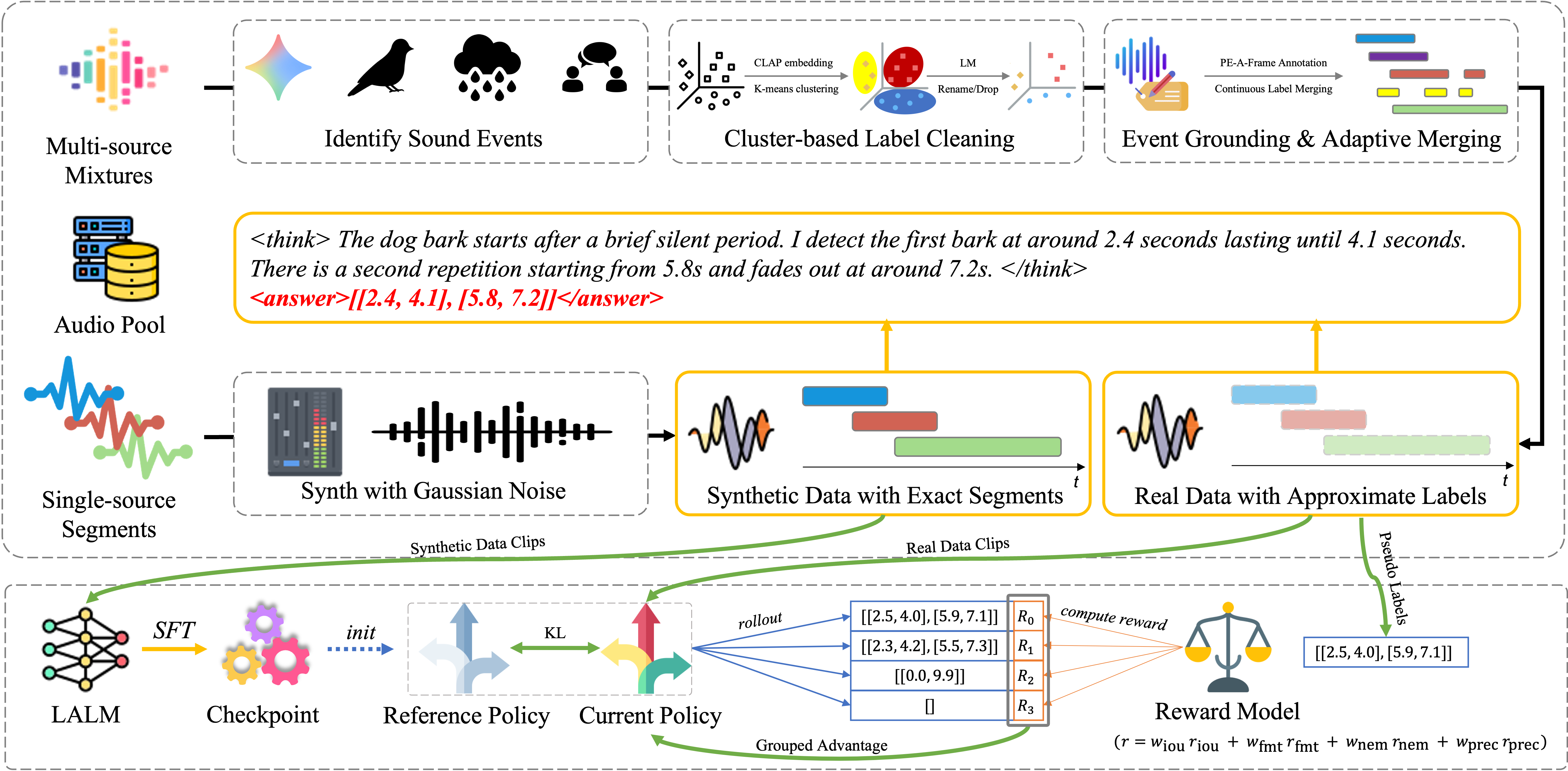}
  \caption{\textbf{Auto-AEG} pipeline for automated data construction and two-stage post-training. The upper block constructs complementary supervision: programmatic mixing with Gaussian background noise yields synthetic clips and placement-exact intervals, while multi-model processing discovers event labels in real recordings, consolidates them through CLAP-based clustering and language-model cleaning, and obtains temporal pseudo-labels with PE-A-Frame and event-dependent merging. Both sources provide query-conditioned interval responses. The lower block uses synthetic supervision for an SFT cold start, then applies GRPO on real recordings. Sampled interval predictions are rewarded for agreement with pseudo-labels, refining temporal localization through the backbone's existing interval-generation interface.}
  \label{fig:teaser}
\end{figure*}

We instantiate this with \textbf{Auto-AEG} (Figure~\ref{fig:teaser}), a scalable automated pipeline 
for constructing open-vocabulary onset/offset supervision. It contains synthesized clips composed programmatically from real audio with placement-exact
ground-truth intervals for an SFT cold-start and real-world recordings paired with pseudo labels constructed through a multi-model collaboration process,
which provide the informative reward signal exploited by the subsequent interval-aware GRPO.
% Synthetic clips, composed programmatically from real audio, carry placement-exact
% ground-truth intervals for an SFT cold-start; real-world clips are pseudo-labeled
% through multi-model collaboration, and provide the informative reward
% signal exploited by Group Relative Policy Optimization (GRPO).

We pair this with \textbf{AEGBench}, an
independently annotated evaluation benchmark drawn from four diverse audio
sources and stratified by hard-case difficulty for diagnostic analysis.

Training Qwen3-Omni-30B-A3B on the automatically constructed data substantially improves temporal
localization on AEGBench and transfers to various held-out benchmarks. The same interval-generation interface supports this transfer without dataset-specific architectural redesign.

Our contributions can be summarized as follows: 

\textbf{(i)} Auto-AEG
, a scalable pipeline that automatically constructs
open-vocabulary onset/offset supervision, providing both placement-exact synthetic data
and in-the-wild pseudo-labels; 

\textbf{(ii)} AEGBench
, an independent quality-filtered benchmark with difficulty stratification
that exposes fine-grained failure modes of audio event temporal grounding;

\textbf{(iii)} a two-stage fine-tuning framework that couples an
instruction-tuned cold start with interval-aware GRPO, which turns
imperfect pseudo-labels into large, architecture-free gains in LALM temporal grounding.

\section{Related Work}

\paragraph{Temporal grounding and sound event detection.}
Video temporal grounding spans proposal-based~\citep{yuan2019semantic}, span-prediction~\citep{zhang2020span}, and transformer-based retrieval~\citep{lei2021detecting,moon2023query} methods. Time-R1~\citep{liu2025timer1comprehensivetemporalreasoning} further demonstrates the value of temporal supervision for timestamp reasoning in video LLMs. In audio, sound event detection (SED), developed through DCASE~\citep{Heittola2020}, uses architectures ranging from CRNNs~\citep{7933050} and PANNs~\citep{10.1109/TASLP.2020.3030497} to AST~\citep{gong2021ast} and BEATs~\citep{chen2022beats}. Weakly supervised methods learn from clip-level labels~\citep{Serizel2018LargeScaleWL,turpault2019sound}. Language-driven detectors extend this setting: DASM~\citep{cai2025detectanysound} performs frame-level retrieval with multimodal queries, while FlexSED~\citep{hai2025flexsed} combines a CLAP text encoder with LLM-assisted query selection. These approaches localize sounds through frame-level detection; our model generates intervals directly as language tokens.

\paragraph{Audio grounding and moment retrieval.}
Text-to-Audio Grounding~\citep{xu2021tag} pairs phrase queries with onset/offset supervision. Audio Moment Retrieval~\citep{munakata2025amr} and CASTELLA~\citep{nakada2025castella} extend this task to real-world, untrimmed recordings. Listening with Time~\citep{shao2026listeningwithtime} provides a 1.2k-hour temporally annotated corpus and evaluates grounding in clips up to 30 minutes. TACOS~\citep{primus2025tacos} supplies 12{,}358 human-annotated Freesound recordings for contrastive retrieval, while composition-based approaches construct timestamp supervision~\citep{xu2024wstag,ahia2025blab,wang2025timeaudio,xie2024audiotime,xie2024picoaudio}.

TEMPO~\citep{kulkarni2026tempo} follows synthetic-to-real SFT with temporal GRPO across speech, sound, and music tasks. It introduces atomic timestamp tokens, a time-aware projector with wall-clock encodings, and a distance-aware Gaussian loss, jointly modifying temporal representation and the training objective. 
Compared to TEMPO, we develop a data-driven route to open-vocabulary event grounding: automatically constructed synthetic placements and real-audio pseudo-labels train the backbone through its existing audio projector and text-token timestamp interface. 
% Its gains arise from constructed supervision and interval-aware optimization without adding a dedicated temporal module or expanding the timestamp vocabulary. The controlled comparison in Section~\ref{sec:ablation} separates GRPO from continued SFT on the same data; Section~\ref{sec:transfer} compares the released systems under shared external evaluation. 
SpotSound~\citep{sun2026spotsoundenhancinglargeaudiolanguage} adapts LALMs using synthetic mixtures and supervised fine-tuning with negative sampling. FineLAP~\citep{li2026finelap} extends CLAP to frame-level alignment, and internal tool use~\citep{an2026framelevel} extracts boundaries from an audio LM's hidden states. We retain native text-token interval prediction and optimize its temporal accuracy through rewards.

\paragraph{Audio language models and reinforcement learning.}
LALMs have advanced auditory understanding~\citep{tang2024salmonn,Qwen-Audio,Qwen2-Audio,huang2025stepaudiounifiedunderstandinggeneration}. Audio Flamingo Next~\citep{ghosh2026audioflamingonextnextgeneration} extends long-audio understanding with intermediate reasoning grounded in timestamps. Our backbone, Qwen3-Omni~\citep{xu2025qwen3omnitechnicalreport}, processes continuous audio features and generates responses autoregressively; targeted temporal supervision addresses the gap between semantic recognition and boundary prediction. Large-scale machine-assisted supervision has precedents in CLIP-based learning~\citep{radford2021learning} and WavCaps caption cleaning~\citep{mei2024wavcaps}. GRPO~\citep{shao2024deepseekmath} provides an alternative to imitating each target sequence by comparing sampled outputs through scalar rewards. We apply this principle to temporal audio grounding, using interval agreement with real-world pseudo-labels as the principal reward signal.

\section{Preliminaries}

\paragraph{Task Definition.}
Let $\mathbf{x} \in \mathbb{R}^T$ be an audio waveform of duration $D$ seconds
and $q$ a natural language query describing a target sound event type.
The model must predict a set of time intervals
$\hat{\mathcal{Y}} = \{[s_i, e_i]\}_{i=1}^{K}$, where each interval
corresponds to an occurrence of the event described by $q$ in $\mathbf{x}$.
Ground-truth segments are denoted $\mathcal{Y}^* = \{[s^*_j, e^*_j]\}_{j=1}^{M}$.
This formulation requires the model to: (a) determine how many occurrences exist
without being given the count at inference time, and (b) generalize over open event vocabulary.

\section{Auto-AEG: Scalable Training Data Construction}
\label{sec:data}
Auto-AEG combines synthetic temporal supervision with real-world pseudo-labels. Synthetic composition provides exact placement times for a supervised cold start; real audio then exposes the model to natural scene structure during interval-aware reinforcement learning. % The two stages use disjoint recording pools.

\subsection{Synthetic Supervision for SFT}
The source pool draws real clips from FSD50K~\citep{fonseca2022fsd50k} and AudioSet~\citep{45857}, which are disjoint from both the AEGBench and the AudioGrounding evaluation split following recording-level deduplication by audio fingerprinting.
Per-frame RMS energy identifies active regions in weakly labeled recordings. We place one or more regions of a target category at scripted onsets and optionally mix in distractors from other categories. Target occurrences remain separate, while distractors may overlap them. Foreground clips are mixed onto a noise background with randomized signal-to-noise ratios, varying acoustic contrast as well as event count and placement. This construction supplies exact placement times while requiring the model to distinguish the queried event from competing sounds. Appendix~\ref{app:synth} gives the sampling distributions and mixing settings.

The synthesizer records every inserted onset and offset. These timestamps are exact with respect to placement, while RMS-derived cuts provide approximate semantic event extents. Real source recordings preserve natural timbre and attack characteristics. Frequent repeated occurrences counter a single-interval response bias; each mixture yields a List-All query and an array of target intervals.

\subsection{Real-World Pseudo-Labels for GRPO}
\paragraph{Event discovery and temporal localization.}
A separate FreeSound pool supplies real recordings for the second stage. Gemini identifies audible events in short audio chunks rather than labeling the entire recording in one pass. Chunk-level inventories are pooled, normalized, and deduplicated; the labeling prompt favors specific, semantically distinct event descriptions over overlapping parent categories. This separates event discovery from boundary estimation and reduces omissions in recordings containing several sound sources.

Each proposed label is classified as continuous or discrete before PE A-Frame~\citep{vyas2025pushingfrontieraudiovisualperception} localizes it through frame-level audio--text similarity. Consecutive active frames form candidate intervals. Short gaps are bridged for continuous sounds, whereas separate detections of discrete sounds retain distinct intervals. Labels without a localized interval are removed. This event-dependent postprocessing suppresses brief dropouts in sustained sounds while preserving repeated transients.

\paragraph{Semantic vocabulary consolidation.}
Labels proposed independently across recordings may contain near-synonyms and redundant descriptions. We embed the unique event names with CLAP~\citep{wu2024largescalecontrastivelanguageaudiopretraining} and group semantically related labels by $k$-means. A language model examines each cluster together with label occurrence counts and assigns keep, drop, or rename decisions. Cluster-level context allows related descriptions to be considered jointly rather than normalized in isolation. The resulting mapping is applied across recordings to consolidate vocabulary while retaining open-text event queries. Clustering groups candidate labels for cleaning; it does not replace them with a fixed set of cluster IDs. Appendix~\ref{app:annot} provides the annotation prompts and temporal postprocessing settings.

% From the annotated candidates, we retain 2,000 clips containing at least two distinct localized event categories, which expand into 5,244 positive clip--category queries. After filtering out queries without a target interval of at least 0.2 seconds, GRPO uses 4,910 queries. 
These real scenes complement the controlled compositions used in SFT, but retain errors from semantic identification and frame-level localization. GRPO uses interval-level rewards to compare sampled responses against this supervision. The continued-SFT control in Section~\ref{sec:ablation} tests whether the benefit follows from the second-stage objective beyond exposure to the same real recordings.

\section{AEGBench: Open-Vocabulary Audio Event Grounding}
\label{sec:benchmark}

To evaluate open-vocabulary audio event grounding, we introduce \textbf{AEGBench}, an independently annotated evaluation benchmark drawn from four diverse audio sources and stratified by hard-case difficulty for diagnostic analysis. AEGBench evaluates temporal grounding on 3,425 real-world clips and 9,924 event queries sampled from AudioSet Strong Labels, FSD50K evaluation audio, BBC Sound Effects, and YouTube life sounds, covering recordings with different collection and labeling conventions.

\subsection{Independent Annotation and Vocabulary}
Every clip is annotated from scratch by three non-author volunteers working independently. Each annotator listens to the audio, identifies audible events, and marks all onset/offset intervals for each event. The AudioSet ontology~\citep{45857} supplies the initial label set. When a sound falls outside this set, annotators propose a new event label, which is introduced by majority agreement among the three annotators. This produces three independently authored sets of event labels and temporal boundaries for every recording. The initial ontology supplies a common semantic reference across sources, while majority-approved additions accommodate sounds beyond its categories.

For matched occurrences, near-identical boundaries with pairwise IoU above $0.95$ are averaged. Occurrences with lower agreement are referred to a senior annotator for listening and adjudication; unresolved ambiguous items are excluded. Appendix~\ref{app:agreement} gives the arbitration protocol. This separation of independent annotation and adjudication makes repeated-event enumeration and uncertain boundaries explicit for benchmark construction.

The training and evaluation vocabularies share AudioSet-related concepts but both allow free-form event names. AudioSet and FSD50K contribute ontology-based categories to the training sources, while Stage~2 proposes and semantically consolidates open-text labels. AEGBench instead expands its initial ontology through human majority decisions. Their common semantic starting point therefore coexists with different label-expansion procedures. Evaluation queries use the resulting human-approved event names through the same text interface as training.

\subsection{Quality and Difficulty Coverage}
Candidate selection uses energy contrast, active ratio, event duration, and per-class caps to retain audible events with identifiable temporal extent. These filters define the evaluation population, including the range represented by its low-contrast subset. Appendix~\ref{app:bench} gives the thresholds and source distributions.

We distinguish target-conditioned difficulty from recording context. \textbf{Polyphonic overlap (PO)}, \textbf{repeated occurrence (RO)}, and \textbf{long duration (LD)} require the queried target itself to overlap another category, recur, or persist. \textbf{Gradual onset/offset (GO)}, \textbf{near silence (NS)}, and \textbf{semantic ambiguity (SA)} retain recording-level tags and characterize the environment in which a query is evaluated. This distinction prevents an unrelated repeated or sustained event from determining the target's subset. Appendix~\ref{app:bench} gives the definitions, source distributions, and subset sizes.

\section{Reward-Guided Fine-Tuning for Temporal Grounding}
\label{sec:training}

We adapt Qwen3-Omni-30B-A3B-Thinking (hereafter referred to as Q3-Omni) with a synthetic SFT cold-start followed by GRPO on real-world pseudo-labels. This design tests whether temporal supervision improves localization on independently annotated AEGBench and whether reward-based optimization extracts more useful signal from noisy labels than continued supervised imitation. Both stages use one prompt template requiring a \texttt{<think>} trace followed by a final \texttt{<answer>}; Appendix~\ref{app:train_config} provides the configuration.

\subsection{Stage 1: SFT Cold-Start}
We train Q3-Omni on synthetic clips using a List-All query that requests every occurrence of a target event. Its supervised \texttt{<think>} scaffold describes the queried event and ground-truth occurrences before the interval answer. Validation loss selects the checkpoint used to initialize GRPO; Appendix~\ref{app:train_config} gives the optimization settings.

\subsection{Stage 2: GRPO Fine-Tuning}
We apply GRPO~\citep{shao2024deepseekmath} to real-world queries with temporal pseudo-labels. AEGBench remains held out from model and reward selection and is used only for final evaluation.

\paragraph{Interval-aware reward.}
The scalar reward combines localization and output-control terms:
\begin{equation}
 r=w_{\mathrm{iou}}r_{\mathrm{iou}}+w_{\mathrm{fmt}}r_{\mathrm{fmt}}
   +w_{\mathrm{nem}}r_{\mathrm{nem}}+w_{\mathrm{prec}}r_{\mathrm{prec}}.
\end{equation}
The localization term $r_{\mathrm{iou}}$ is F1-IoU@0.5, the harmonic mean of Recall-IoU@0.5 and Precision-IoU@0.5. Balancing missed and spurious intervals discourages dense interval predictions. The format term rewards a valid \texttt{<think>}/\texttt{<answer>} structure and JSON array; the non-empty term discourages empty predictions; and the count-based precision term penalizes excessive segmentation.

Every retained training query has a reference interval, so these terms can be evaluated consistently. Together they encourage complete, parseable answers while limiting spurious events. Appendix~\ref{app:reward_settings} gives the exact reward implementation.

For sampled response $i$, GRPO uses the group-normalized advantage
$$\mathrm{adv}_i=(r_i-\bar r)/(\mathrm{std}(r)+\varepsilon)$$
An L2 penalty between policy LoRA weights and the frozen SFT adapter anchors optimization to the cold-start initialization. The regularization coefficient and full optimization settings appear in Appendix~\ref{app:train_config}; Section~\ref{sec:ablation} evaluates the contribution of the two stages.

\begin{table*}[t]
\centering\footnotesize
\setlength{\tabcolsep}{2pt}
\begin{tabularx}{\linewidth}{@{}l*{12}{C}@{}}
\toprule
& \multicolumn{6}{c}{\textbf{Overall performance}} & \multicolumn{6}{c}{\textbf{Difficulty-subset mIoU}}\\
\cmidrule(lr){2-7}\cmidrule(lr){8-13}
\textbf{Model} & \textbf{mIoU} & \textbf{IF1} & \textbf{SF1} & \textbf{IP} & \textbf{IR} & \textbf{R@.5} & \textbf{PO} & \textbf{GO} & \textbf{RO} & \textbf{NS} & \textbf{SA} & \textbf{LD}\\
\midrule
\multicolumn{13}{l}{\emph{External zero-shot LALMs}}\\
Gemini-3-Pro & 0.324 & 0.224 & 0.576 & 0.236 & 0.214 & 0.291 & 0.298 & 0.328 & 0.191 & 0.332 & 0.268 & 0.388 \\
Kimi-Audio-7B & 0.117 & 0.018 & 0.281 & 0.010 & 0.077 & 0.070 & 0.114 & 0.129 & 0.130 & 0.108 & 0.115 & 0.073 \\
Qwen2-Audio-7B & 0.157 & 0.068 & 0.349 & 0.063 & 0.075 & 0.083 & 0.147 & 0.167 & 0.133 & 0.154 & 0.140 & 0.042 \\
Audio Flamingo 3 & 0.233 & 0.046 & 0.533 & 0.027 & 0.162 & 0.168 & 0.220 & 0.237 & 0.227 & 0.237 & 0.204 & 0.129 \\
\midrule
\multicolumn{13}{l}{\emph{Dedicated grounding models}}\\
CLAP & 0.303 & 0.203 & 0.566 & 0.275 & 0.160 & 0.263 & 0.282 & 0.310 & 0.113 & 0.302 & 0.257 & \textbf{0.574} \\
TAG & 0.243 & 0.166 & 0.487 & 0.129 & 0.231 & 0.208 & 0.221 & 0.248 & 0.289 & 0.252 & 0.205 & 0.166 \\
DASM & 0.206 & 0.270 & 0.280 & \textbf{0.443} & 0.194 & 0.206 & 0.184 & 0.209 & 0.207 & 0.223 & 0.154 & 0.165 \\
PE-A-Frame & 0.392 & 0.219 & 0.610 & 0.160 & 0.347 & 0.374 & 0.359 & 0.400 & \textbf{0.373} & 0.404 & 0.301 & 0.438 \\
\midrule
\multicolumn{13}{l}{\emph{Task-adapted LALMs}}\\
TEMPO SFT & 0.383 & 0.194 & 0.639 & 0.142 & 0.307 & 0.360 & 0.351 & 0.385 & 0.292 & 0.391 & 0.306 & 0.349 \\
\quad + GRPO & 0.425 & 0.271 & 0.665 & 0.224 & 0.344 & 0.410 & 0.392 & 0.428 & 0.317 & 0.432 & 0.346 & 0.449 \\
\midrule
\multicolumn{13}{l}{\emph{Auto-AEG adaptation}}\\
Qwen3-Omni & 0.278 & 0.197 & 0.530 & 0.214 & 0.183 & 0.246 & 0.252 & 0.286 & 0.158 & 0.287 & 0.224 & 0.397 \\
\quad + SFT & 0.372 & 0.209 & 0.644 & 0.201 & 0.218 & 0.318 & 0.350 & 0.387 & 0.188 & 0.337 & 0.340 & 0.149 \\
\quad + SFT + GRPO & \textbf{0.480} & \textbf{0.402} & \textbf{0.696} & 0.421 & \textbf{0.384} & \textbf{0.465} & \textbf{0.444} & \textbf{0.490} & 0.343 & \textbf{0.483} & \textbf{0.402} & 0.545 \\
\bottomrule
\end{tabularx}
\caption{AEGBench results on the complete 9,924-query evaluation set. IF1/IP/IR denote interval F1/precision/recall under one-to-one matching at IoU $\geq0.5$, with micro aggregation. SF1 is one-second Segment F1; R@.5 is best-match Recall-IoU@0.5. mIoU, SF1, and R@.5 are query-macro. PO/RO/LD concern the queried target; GO/NS/SA describe recording context. Bold marks the best score per column.}
\label{tab:main_results}\label{tab:hardcase_full}
\end{table*}

\section{Experiments}
\label{sec:experiments}

\subsection{Evaluation Setup}
We evaluate Qwen3-Omni-30B-A3B before adaptation, after Stage~1 SFT, and after Stage~2 GRPO on the held-out AEGBench queries. All LALMs receive the same List-All query, and evaluation uses only the intervals extracted from their responses. Each model receives the full audio through its native interface; model-specific handling and inference settings appear in Appendix~\ref{app:audio_length} and~\ref{app:train_config}.

\paragraph{Baselines.}
We compare with four external zero-shot LALMs: Gemini-3-Pro, Kimi-Audio-7B-Instruct~\citep{kimiaudio2025}, Qwen2-Audio-7B-Instruct~\citep{Qwen2-Audio}, and Audio Flamingo 3~\citep{goel2025audioflamingo3advancing}. Dedicated grounding models include TAG~\citep{xu2021tag,xu2024wstag}, sliding-window CLAP~\citep{wu2024largescalecontrastivelanguageaudiopretraining}, DASM~\citep{cai2025detectanysound} and PE-A-Frame-Large~\citep{vyas2025pushingfrontieraudiovisualperception}. The task-adapted comparison includes TEMPO SFT and GRPO~\citep{kulkarni2026tempo}. This comparison covers general audio-language models, specialized detectors, and task-adapted LALMs.

\paragraph{Metrics.}
mIoU averages each reference interval's best predicted overlap, and Recall-IoU@$\theta$ (R@$\theta$) measures recovery at IoU $\geq\theta$. All primary comparisons use Interval F1@0.5 with one-to-one matching, together with interval precision/recall where reported. Interval F1/P/R are micro-averaged; mIoU, R@0.5, and one-second Segment F1 are query-macro. All AEGBench model comparisons use the complete released query set and full recording durations, without an audio-length cutoff. Appendix~\ref{app:metrics} gives the definitions; onset metrics remain additional boundary diagnostics.

\subsection{Main Results and Baseline Comparison}
Table~\ref{tab:main_results} shows distinct effects of the two training stages. Synthetic SFT improves temporal coverage and interval F1, but its lower interval precision indicates that recovering more activity does not yet calibrate candidate selection. GRPO raises IF1 from $0.209$ to $0.402$, alongside improved coverage and precision. These combined gains indicate more accurate instance recovery and fewer spurious detections.

\paragraph{Task-adapted models.}
TEMPO improves overlap, interval F1, precision, and recall from SFT to GRPO, supporting temporal reward optimization on another backbone. Both checkpoints improve substantially over Audio Flamingo 3. Auto-AEG exceeds TEMPO GRPO by $0.055$ mIoU, alongside stronger segment agreement and IF1. Both recipes benefit from temporal adaptation, while their training resources and representations differ.

\paragraph{General models and dedicated detectors.}
PE-A-Frame has stronger overlap than the general zero-shot audio-language models, but its lower interval precision and F1 show that broad coverage can coexist with extra or fragmented detections. DASM has the highest interval precision yet lower recall and overlap than Auto-AEG: selecting fewer segments improves precision but misses requested occurrences. Auto-AEG combines the strongest interval F1 with broad recovery. Reporting overlap, one-to-one matching, and temporal activity agreement together distinguishes these prediction behaviors.

\subsection{Difficulty-Stratified Analysis}
Table~\ref{tab:hardcase_full} separates target-conditioned PO/RO/LD from recording-context GO/NS/SA. Auto-AEG exceeds both TEMPO checkpoints in overlap and recovery on all three target-conditioned subsets. Within Qwen3-Omni, GRPO also improves these subsets over SFT, including sustained targets for which SFT falls below the base model.

The ranking depends on the localization criterion. PE-A-Frame achieves higher overlap and recall on repeated targets, and CLAP has stronger coverage on sustained targets. Auto-AEG nevertheless has the highest IF1 on all three target-conditioned subsets (Appendix~\ref{app:detectors}); on RO, its $0.366$ exceeds PE-A-Frame's $0.241$, indicating more reliable one-to-one instance recovery. A broad span can overlap several repeated events without recovering them as separate instances; one-to-one matching makes this distinction visible. The GO/NS/SA results describe performance in recordings containing gradual transitions, weak activity, or ambiguous semantics; they do not isolate those properties of every queried target.

\subsection{Cross-Dataset Generalization}
\label{sec:transfer}
\paragraph{AudioGrounding.}
AudioGrounding~\citep{xu2021tag} uses short AudioCaps clips and free-text event phrases extracted from human-written captions, providing a complementary evaluation distribution. All models receive the full clip and use the same interval-scoring pipeline. Table~\ref{tab:audiogrounding} shows monotonic improvement across Qwen3-Omni stages in coverage, segment agreement, and strict interval F1. The GRPO gain therefore extends beyond synthetic training compositions to independently annotated recordings. TEMPO also improves overlap and interval F1 from SFT to GRPO and substantially exceeds its Audio Flamingo 3 base. Both adaptation recipes yield gains beyond their respective unadapted models.

TEMPO GRPO and Auto-AEG have similar SF1 (both $0.842$), yet Auto-AEG's IF1 is $0.131$ higher, showing the gap between covering active time and reconstructing individual occurrences.

\begin{table}[t]
\centering\small
\setlength{\tabcolsep}{3pt}
\begin{tabular}{@{}lrrrr@{}}
\toprule
\textbf{Model} & \textbf{mIoU} & \textbf{R@0.5} & \textbf{SF1} & \textbf{IF1}\\
\midrule
\multicolumn{5}{l}{\emph{External zero-shot LALMs}}\\
Gemini-3-Pro & 0.465 & 0.457 & 0.753 & 0.400 \\
Kimi-Audio-7B & 0.143 & 0.103 & 0.363 & 0.036 \\
Qwen2-Audio-7B & 0.248 & 0.152 & 0.557 & 0.127 \\
Audio Flamingo 3 & 0.407 & 0.367 & 0.732 & 0.140 \\
\midrule
\multicolumn{5}{l}{\emph{Dedicated grounding models}}\\
CLAP & 0.182 & 0.168 & 0.309 & 0.163 \\
TAG & 0.377 & 0.346 & 0.713 & 0.224 \\
DASM & 0.316 & 0.320 & 0.447 & 0.411 \\
PE-A-Frame & 0.538 & 0.556 & 0.771 & 0.315 \\
\midrule
\multicolumn{5}{l}{\emph{Task-adapted LALMs}}\\
TEMPO SFT & 0.586 & 0.594 & 0.833 & 0.388 \\
\quad + GRPO & 0.617 & 0.636 & 0.842 & 0.492 \\
\midrule
\multicolumn{5}{l}{\emph{Auto-AEG adaptation}}\\
Qwen3-Omni & 0.364 & 0.370 & 0.605 & 0.342 \\
\quad + SFT & 0.479 & 0.493 & 0.748 & 0.465 \\
\quad + SFT + GRPO & \textbf{0.626} & \textbf{0.658} & \textbf{0.842} & \textbf{0.623} \\
\bottomrule
\end{tabular}
\caption{AudioGrounding results on the complete 997-query evaluation set. Interval F1@0.5 is micro-averaged; the other metrics are query-macro.}
\label{tab:audiogrounding}
\end{table}

\paragraph{DESED.}
% We evaluate transfer on the official DESED~\citep{Serizel2018LargeScaleWL} public evaluation recordings, using the same positive class queries and strong temporal annotations for the external baselines and all three Qwen3-Omni checkpoints. Table~\ref{tab:desed} reports query-macro mIoU and micro-averaged event F1 and precision under one-to-one matching at IoU $\geq0.5$. For a common operating-point comparison, every final predicted interval receives confidence $1.0$ when computing PSDS~\citep{bilen2020framework}. 
% PSDS evaluates event coverage through intersection-based detection and ground-truth criteria, whereas event F1 requires one-to-one interval matching at $\text{IoU}\geq0.5$. These criteria respond differently to merged detections and fragmented coverage, particularly under the looser PSDS$_2$ setting. %Here, PSDS is computed from a single operating point, complementing instance-level evaluation with coverage-based detection performance at the selected output setting.

We evaluate transfer on the DESED~\citep{Serizel2018LargeScaleWL} public evaluation recordings, using identical positive class queries and strong temporal annotations. Table~\ref{tab:desed} reports query-macro mIoU and micro-averaged event F1 and precision with one-to-one matching at IoU $\geq 0.5$. PSDS~\citep{bilen2020framework} is computed at a single operating point by assigning confidence $1.0$ to every predicted interval. Its intersection-based criteria assess coverage differently from one-to-one IoU matching, particularly under the looser PSDS$_2$ setting.

PE-A-Frame achieves the highest mIoU but lower event precision and F1, while DASM has the highest event precision but lower mIoU. Auto-AEG achieves higher event F1 than both detectors while maintaining broad temporal coverage. TEMPO's GRPO stage improves event F1 and precision over its SFT checkpoint, with a smaller gain in mIoU. Auto-AEG outperforms TEMPO's released GRPO model in both mIoU and event F1. This comparison reflects the complete training recipes, given differences in backbone, temporal representation, training data, and task scope.

Within Qwen3-Omni, SFT improves mIoU and segment agreement but lowers event precision and F1, suggesting improved coverage without reliable separation of individual occurrences. Interval-aware GRPO improves coverage, event F1, and precision together.
CLAP's strong PSDS$_2$ and weaker interval F1 show how intersection-based activity coverage can diverge from one-to-one event recovery, motivating both measures.

\begin{table}[t]
\centering\footnotesize
\setlength{\tabcolsep}{2.5pt}
\begin{tabular}{@{}lrrrrr@{}}
\toprule
\textbf{Model} & \textbf{PSDS$_1$} & \textbf{PSDS$_2$} & \textbf{mIoU} & \textbf{IF1} & \textbf{IP}\\
\midrule
\multicolumn{6}{l}{\emph{External zero-shot LALMs}}\\
Gemini-3-Pro & 0.360 & 0.688 & 0.507 & 0.347 & 0.440\\
Kimi-Audio-7B & 0.159 & 0.309 & 0.184 & 0.124 & 0.099\\
Qwen2-Audio-7B & 0.214 & 0.306 & 0.419 & 0.227 & 0.216\\
Audio Flamingo 3 & 0.135 & 0.218 & 0.492 & 0.197 & 0.139\\
\midrule
\multicolumn{6}{l}{\emph{Dedicated grounding models}}\\
CLAP & 0.316 & 0.798 & 0.375 & 0.221 & 0.396\\
TAG & 0.211 & 0.334 & 0.456 & 0.409 & 0.335\\
DASM & 0.143 & 0.178 & 0.392 & 0.566 & \textbf{0.805}\\
PE-A-Frame & 0.163 & 0.177 & \textbf{0.727} & 0.476 & 0.395\\
\midrule
\multicolumn{6}{l}{\emph{Task-adapted LALMs}}\\
TEMPO SFT & 0.373 & 0.539 & 0.653 & 0.510 & 0.511\\
\quad + GRPO & 0.395 & 0.583 & 0.658 & 0.540 & 0.576\\
\midrule
\multicolumn{6}{l}{\emph{Auto-AEG adaptation}}\\
Qwen3-Omni & 0.306 & 0.556 & 0.392 & 0.302 & 0.454\\
\quad + SFT & 0.305 & 0.598 & 0.416 & 0.264 & 0.357\\
\quad + SFT + GRPO & \textbf{0.591} & \textbf{0.819} & 0.673 & \textbf{0.590} & 0.739\\
\bottomrule
\end{tabular}
\caption{Cross-dataset results on DESED public evaluation (1,112 positive queries). All rows use the same recordings and unified scoring code. PSDS uses one confidence operating point; mIoU is query-macro, while event F1 and precision are micro-averaged.}% Bold marks the best score per column.}
\label{tab:desed}\label{tab:desed_full}
\end{table}

% PE-A-Frame leads overlap but has lower event precision and F1, suggesting extra or fragmented detections. DASM has the strongest event precision, yet its weaker query-level overlap reveals incomplete recovery. Auto-AEG combines broader coverage with stronger event F1 than both detectors, balancing instance recovery and selectivity.

% TEMPO provides a closer methodological comparison. Its GRPO stage improves one-to-one event matching and precision over its own SFT checkpoint, with a smaller change in average overlap. Auto-AEG achieves stronger overlap and event matching than TEMPO's released GRPO model on the same DESED queries, while TEMPO's SFT checkpoint already provides substantial interval coverage. The models differ in backbone, temporal representation, training data, and task scope, so these results compare their complete training recipes rather than attributing the gap to GRPO alone.

% Within Qwen3-Omni, SFT improves average overlap and segment agreement but reduces event precision and F1, suggesting that broader coverage does not yet separate individual occurrences reliably. Interval-aware GRPO improves coverage, event matching, and event precision together. 

\subsection{Training Objective and Reward Ablations}
\label{sec:ablation}
\paragraph{GRPO versus continued SFT.}
We replace Stage~2 GRPO with continued SFT on the same real-audio query pool and Stage~1 initialization, keeping the source of temporal supervision fixed. Table~\ref{tab:ablation_grpo} shows that continued SFT increases recall but reduces IF1 to $0.183$, versus $0.402$ for GRPO; overlap remains nearly unchanged and segment agreement decreases. GRPO instead improves overlap, segment agreement, and interval F1 together, producing a more consistent localization gain.

Continued SFT maximizes the likelihood of a serialized reference response. GRPO compares sampled candidates through interval agreement, favoring accurate temporal predictions even when event labels and output formats match. The control supports interval-aware optimization over additional supervised imitation on this noisy real-data corpus. The reward ablations below examine the localization signal and auxiliary output constraints.

\begin{table}[t]
\centering\small
\setlength{\tabcolsep}{4pt}
\begin{tabular}{@{}lrrrr@{}}
\toprule
\textbf{Training} & \textbf{mIoU} & \textbf{IF1} & \textbf{SF1} & \textbf{R@0.5}\\
\midrule
Stage 1 SFT & 0.372 & 0.209 & 0.644 & 0.318 \\
+ real-data SFT & 0.371 & 0.183 & 0.620 & 0.348 \\
+ real-data GRPO & \textbf{0.480} & \textbf{0.402} & \textbf{0.696} & \textbf{0.465} \\
\bottomrule
\end{tabular}
\caption{Objective ablation on AEGBench under the same interval-scoring protocol as Table~\ref{tab:main_results}. Both second-stage variants share the real-data query pool and SFT initialization; F1 uses interval matching and micro aggregation.}
\label{tab:ablation_grpo}
\end{table}

\paragraph{Reward components.}
Table~\ref{tab:reward_components} isolates each reward term while holding the SFT initialization and training queries fixed. Removing localization causes the largest loss in overlap and recovery, establishing interval agreement as the main optimization signal. Removing the auxiliary count-based term increases recall but reduces onset precision: accepting more candidate segments recovers additional occurrences at the cost of less selective starts. This term controls excessive segmentation rather than directly penalizing long intervals. Format and non-empty rewards have smaller effects on overlap, supporting their role as output constraints alongside the localization objective. Additional onset metrics for the objective comparison appear in Appendix~\ref{app:ablation_onset}.

\begin{table}[t]
\centering\small
\setlength{\tabcolsep}{4pt}
\begin{tabular}{@{}lrrr@{}}
\toprule
\textbf{Reward setting} & \textbf{mIoU} & \textbf{R@0.5} & \textbf{OP}\\
\midrule
Full reward & \textbf{0.480} & 0.465 & \textbf{0.559}\\
$w_{\mathrm{iou}}=0$ & 0.388 & 0.337 & 0.501\\
$w_{\mathrm{fmt}}=0$ & 0.473 & 0.459 & 0.549\\
$w_{\mathrm{nem}}=0$ & 0.476 & 0.462 & 0.556\\
$w_{\mathrm{prec}}=0$ & 0.469 & \textbf{0.473} & 0.497\\
\bottomrule
\end{tabular}
\caption{Reward-component ablation on AEGBench.
%  Each variant sets one weight to zero without changing the other terms. 
OP denotes onset precision with a $0.5$\,s collar.}
\label{tab:reward_components}
\end{table}

% A separate Qwen2.5-Omni-7B comparison shows that the default reward weights are not uniformly best: removing the auxiliary precision term improves overlap and strict event matching over standard-reward GRPO on DESED and an audio-length-limited AEGBench subset (Appendix~\ref{app:reward_7b}). The localization reward still includes precision through F1-IoU; this comparison changes only the additional count-based term.

\section{Conclusion}
We present Auto-AEG, an automated pipeline for constructing open-vocabulary temporal audio supervision. It combines programmatic synthesis with multi-model annotation of real recordings to provide scalable training data with event labels and onset--offset intervals. Two-stage SFT and interval-aware GRPO demonstrate the value of this data for improving temporal localization and transferring to external benchmarks. We also introduce AEGBench, an independently annotated benchmark with difficulty subsets for evaluating event recovery and boundary alignment. Together, these contributions establish automatic data construction as a practical route to improving temporal grounding in existing audio-language models.

\section*{Limitations}
% Our main comparison uses one Qwen3-Omni backbone and one matched continued-SFT configuration. Additional models and seeds would test the stability of these findings. 
The Auto-AEG data construction pipeline uses Gemini-3-Pro/Gemini-2.5-Flash for label post-processing and PE-A-Frame for coarse localization. This reliance on pretrained models may introduce model-dependent biases, while PE-A-Frame's native 40 ms temporal resolution limits annotation granularity. Synthetic labels record exact placements of RMS-selected source regions, so their semantic extent can still be approximate. Real-world training labels inherit recognition and boundary errors from the teacher models, which may produce systematic errors in the reward.

AEGBench samples four audio sources and applies salience and active-ratio filters, so its difficulty subsets cover the retained evaluation range. Human majority decisions expand the initial AudioSet vocabulary, but previously unseen query compositions require further evaluation. Cross-dataset comparisons also depend on the backbones and training resources used. 
% DESED PSDS reflects one operating point rather than a threshold sweep. 

% \section*{Ethical considerations}
% Auto-AEG is intended for research on locating named sound events in recordings. Such localization can support audio search and accessibility, but it could also be used to monitor recordings without the knowledge of the people captured in them. Source audio may contain incidental speech or other identifying sounds; our labels describe audible events rather than speaker identities or personal attributes. Synthetic mixtures and model-generated pseudo-labels can carry errors and source-specific biases, while human annotation does not eliminate ambiguity at event boundaries. We therefore do not treat model predictions as evidence about individuals or as a sole basis for safety-critical decisions. We plan to share annotation files only, without redistributing third-party raw audio; use of the source recordings remains subject to their original terms.

\bibliography{custom}

\begin{thebibliography}{46}
\providecommand{\natexlab}[1]{#1}

\bibitem[{Ahia et~al.(2025)Ahia, Bartelds et~al.}]{ahia2025blab}
Orevaoghene Ahia, Martijn Bartelds, et~al. 2025.
\newblock {BLAB}: Brutally long audio bench.
\newblock \emph{arXiv preprint arXiv:2505.03054}.

\bibitem[{An et~al.(2026)An, Keung, Wang, Ahia, and Smith}]{an2026framelevel}
Joesph An, Phillip Keung, Jiaqi Wang, Orevaoghene Ahia, and Noah~A. Smith. 2026.
\newblock \href {https://arxiv.org/abs/2602.10230} {Frame-level internal tool use for temporal grounding in audio {LMs}}.
\newblock \emph{Preprint}, arXiv:2602.10230.

\bibitem[{Bilen et~al.(2020)Bilen, Ferroni, Tuveri, Azcarreta, and Krstulovi{\'c}}]{bilen2020framework}
{\c{C}}a{\u{g}}da{\c{s}} Bilen, Giacomo Ferroni, Francesco Tuveri, Juan Azcarreta, and Sacha Krstulovi{\'c}. 2020.
\newblock A framework for the robust evaluation of sound event detection.
\newblock In \emph{ICASSP 2020-2020 IEEE International Conference on Acoustics, Speech and Signal Processing (ICASSP)}, pages 61--65. IEEE.

\bibitem[{Cai et~al.(2025)Cai, Song, Gu, Jiang, Song, and McLoughlin}]{cai2025detectanysound}
Pengfei Cai, Yan Song, Qing Gu, Nan Jiang, Haoyu Song, and Ian McLoughlin. 2025.
\newblock \href {https://arxiv.org/abs/2507.16343} {{Detect Any Sound}: Open-vocabulary sound event detection with multi-modal queries}.
\newblock \emph{Preprint}, arXiv:2507.16343.

\bibitem[{{\c C}ak{\i}r et~al.(2017){\c C}ak{\i}r, Parascandolo, Heittola, Huttunen, and Virtanen}]{7933050}
Emre {\c C}ak{\i}r, Giambattista Parascandolo, Toni Heittola, Heikki Huttunen, and Tuomas Virtanen. 2017.
\newblock \href {https://doi.org/10.1109/TASLP.2017.2690575} {Convolutional recurrent neural networks for polyphonic sound event detection}.
\newblock \emph{IEEE/ACM Transactions on Audio, Speech, and Language Processing}, 25(6):1291--1303.

\bibitem[{Chen et~al.(2023)Chen, Wu, Wang, Liu, Tompkins, Chen, and Wei}]{chen2022beats}
Sanyuan Chen, Yu~Wu, Chengyi Wang, Shujie Liu, Daniel Tompkins, Zhuo Chen, and Furu Wei. 2023.
\newblock {BEATs}: Audio pre-training with acoustic tokenizers.
\newblock In \emph{International Conference on Machine Learning (ICML)}.

\bibitem[{Chu et~al.(2024)Chu, Xu, Yang, Wei, Wei, Guo, Leng, Lv, He, Lin, Zhou, and Zhou}]{Qwen2-Audio}
Yunfei Chu, Jin Xu, Qian Yang, Haojie Wei, Xipin Wei, Zhifang Guo, Yichong Leng, Yuanjun Lv, Jinzheng He, Junyang Lin, Chang Zhou, and Jingren Zhou. 2024.
\newblock Qwen2-audio technical report.
\newblock \emph{arXiv preprint arXiv:2407.10759}.

\bibitem[{Chu et~al.(2023)Chu, Xu, Zhou, Yang, Zhang, Yan, Zhou, and Zhou}]{Qwen-Audio}
Yunfei Chu, Jin Xu, Xiaohuan Zhou, Qian Yang, Shiliang Zhang, Zhijie Yan, Chang Zhou, and Jingren Zhou. 2023.
\newblock Qwen-audio: Advancing universal audio understanding via unified large-scale audio-language models.
\newblock \emph{arXiv preprint arXiv:2311.07919}.

\bibitem[{Deshmukh et~al.(2023)Deshmukh, Elizalde, Singh, and Wang}]{deshmukh2023pengi}
Soham Deshmukh, Benjamin Elizalde, Rita Singh, and Huaming Wang. 2023.
\newblock Pengi: An audio language model for audio tasks.
\newblock In \emph{Advances in Neural Information Processing Systems}, volume~36, pages 18090--18108.

\bibitem[{Fonseca et~al.(2022)Fonseca, Favory, Pons, Font, and Serra}]{fonseca2022fsd50k}
Eduardo Fonseca, Xavier Favory, Jordi Pons, Frederic Font, and Xavier Serra. 2022.
\newblock {FSD50K}: An open dataset of human-labeled sound events.
\newblock \emph{IEEE/ACM Transactions on Audio, Speech, and Language Processing}, 30:829--852.

\bibitem[{Gemmeke et~al.(2017)Gemmeke, Ellis, Freedman, Jansen, Lawrence, Moore, Plakal, and Ritter}]{45857}
Jort~F. Gemmeke, Daniel P.~W. Ellis, Dylan Freedman, Aren Jansen, Wade Lawrence, R.~Channing Moore, Manoj Plakal, and Marvin Ritter. 2017.
\newblock Audio set: An ontology and human-labeled dataset for audio events.
\newblock In \emph{Proc. IEEE ICASSP 2017}, New Orleans, LA.

\bibitem[{Ghosh et~al.(2026)Ghosh, Goel, Jayakumar, Koroshinadze, Anand, Kong, Gururani, gil Lee, Kim, Aljafari, Yang, Kim, Duraiswami, Manocha, Shoeybi, Catanzaro, Liu, and Ping}]{ghosh2026audioflamingonextnextgeneration}
Sreyan Ghosh, Arushi Goel, Kaousheik Jayakumar, Lasha Koroshinadze, Nishit Anand, Zhifeng Kong, Siddharth Gururani, Sang gil Lee, Jaehyeon Kim, Aya Aljafari, Chao-Han~Huck Yang, Sungwon Kim, Ramani Duraiswami, Dinesh Manocha, Mohammad Shoeybi, Bryan Catanzaro, Ming-Yu Liu, and Wei Ping. 2026.
\newblock \href {https://arxiv.org/abs/2604.10905} {Audio flamingo next: Next-generation open audio-language models for speech, sound, and music}.
\newblock \emph{Preprint}, arXiv:2604.10905.

\bibitem[{Goel et~al.(2025)Goel, Ghosh, Kim, Kumar, Kong, gil Lee, Yang, Duraiswami, Manocha, Valle, and Catanzaro}]{goel2025audioflamingo3advancing}
Arushi Goel, Sreyan Ghosh, Jaehyeon Kim, Sonal Kumar, Zhifeng Kong, Sang gil Lee, Chao-Han~Huck Yang, Ramani Duraiswami, Dinesh Manocha, Rafael Valle, and Bryan Catanzaro. 2025.
\newblock \href {https://arxiv.org/abs/2507.08128} {Audio flamingo 3: Advancing audio intelligence with fully open large audio language models}.
\newblock \emph{Preprint}, arXiv:2507.08128.

\bibitem[{Gong et~al.(2021)Gong, Chung, and Glass}]{gong2021ast}
Yuan Gong, Yu-An Chung, and James Glass. 2021.
\newblock Ast: Audio spectrogram transformer.
\newblock \emph{arXiv preprint arXiv:2104.01778}.

\bibitem[{Gong et~al.(2024)Gong, Luo, Liu, Karlinsky, and Glass}]{gong2024ltu}
Yuan Gong, Hongyin Luo, Alexander~H. Liu, Leonid Karlinsky, and James Glass. 2024.
\newblock Listen, think, and understand.
\newblock In \emph{International Conference on Learning Representations}.

\bibitem[{Hai et~al.(2025)Hai, Wang, Guo, and Elhilali}]{hai2025flexsed}
Jiarui Hai, Helin Wang, Weizhe Guo, and Mounya Elhilali. 2025.
\newblock \href {https://arxiv.org/abs/2509.18606} {{FlexSED}: Towards open-vocabulary sound event detection}.
\newblock \emph{Preprint}, arXiv:2509.18606.

\bibitem[{Heittola et~al.(2020)Heittola, Mesaros, and Virtanen}]{Heittola2020}
Toni Heittola, Annamaria Mesaros, and Tuomas Virtanen. 2020.
\newblock \href {https://arxiv.org/abs/2005.14623} {Acoustic scene classification in dcase 2020 challenge: generalization across devices and low complexity solutions}.
\newblock In \emph{Proceedings of the Detection and Classification of Acoustic Scenes and Events 2020 Workshop (DCASE2020)}, pages 56--60.

\bibitem[{Huang et~al.(2025)Huang, Wu, Wang, Yan, Hu, Feng, Tian, Shen, Li, Chen, Liu, Miao, You, Chen, Yang, Huang, Zhang, Gong, Zhang, Zhou, Sun, Li, Feng, Wan, Hu, Wu, Zhen, Ming, Yuan, Zhang, Zhou, Li, Ma, Wang, An, Ji, Li, Wen, Kong, Ma, Liang, Mou, Ahmidi, Wang, Li, Miao, Xu, Wang, Shi, Sun, Hu, Sai, Liu, Huang, Yan, Wang, Jia, Zhang, Gong, Guo, Liu, Liu, Feng, Wu, Wu, Yang, Wang, Zhang, Lin, Li, Xia, Zhou, Zhao, Gu, Chen, Wu, Li, Li, Li, Liang, Wang, Hao, Wu, Tan, Sun, Shuai, Pang, Yang, Gao, Yuan, Liu, Deng, Jiang, Liu, Cao, Wang, Deng, Xie, Ming, He, Sun, Han, Huang, Deng, Liu, Wu, Zhao, Wei, Yu, Cao, Li, Ma, Xu, Wang, Shi, Wang, Zhou, Zhong, Zhang, Wei, Luo, Lu, Yin, Luo, Ding, Yan, Dai, Yang, Xie, Ge, Sun, Huang, Chang, Guan, Yang, Zhang, Jiao, Jiang, Shum, Chen, Li, Zhou, Zhang, Zhang, and Zhu}]{huang2025stepaudiounifiedunderstandinggeneration}
Ailin Huang, Boyong Wu, Bruce Wang, Chao Yan, Chen Hu, Chengli Feng, Fei Tian, Feiyu Shen, Jingbei Li, Mingrui Chen, Peng Liu, Ruihang Miao, Wang You, Xi~Chen, Xuerui Yang, Yechang Huang, Yuxiang Zhang, Zheng Gong, Zixin Zhang, and 126 others. 2025.
\newblock \href {https://arxiv.org/abs/2502.11946} {Step-audio: Unified understanding and generation in intelligent speech interaction}.
\newblock \emph{Preprint}, arXiv:2502.11946.

\bibitem[{{Kimi Team}(2025)}]{kimiaudio2025}
{Kimi Team}. 2025.
\newblock Kimi-audio technical report.
\newblock \url{https://github.com/MoonshotAI/Kimi-Audio}.
\newblock Open-source 7B audio language model with native temporal grounding support.

\bibitem[{Kong et~al.(2020)Kong, Cao, Iqbal, Wang, Wang, and Plumbley}]{10.1109/TASLP.2020.3030497}
Qiuqiang Kong, Yin Cao, Turab Iqbal, Yuxuan Wang, Wenwu Wang, and Mark~D. Plumbley. 2020.
\newblock \href {https://doi.org/10.1109/TASLP.2020.3030497} {Panns: Large-scale pretrained audio neural networks for audio pattern recognition}.
\newblock \emph{IEEE/ACM Trans. Audio, Speech and Lang. Proc.}, 28:2880–2894.

\bibitem[{Kulkarni et~al.(2026)Kulkarni, Jayakumar, Ghosh, Aich, Duraiswami, and Manocha}]{kulkarni2026tempo}
Apoorva Kulkarni, Kaousheik Jayakumar, Sreyan Ghosh, Utathya Aich, Ramani Duraiswami, and Dinesh Manocha. 2026.
\newblock \href {https://arxiv.org/abs/2608.29999} {{TEMPO}: Temporally-grounded multi-task post-training for large audio-language models}.
\newblock \emph{Preprint}, arXiv:2608.29999.

\bibitem[{Lei et~al.(2021)Lei, Berg, and Bansal}]{lei2021detecting}
Jie Lei, Tamara~L Berg, and Mohit Bansal. 2021.
\newblock Detecting moments and highlights in videos via natural language queries.
\newblock \emph{Advances in Neural Information Processing Systems}, 34:11846--11858.

\bibitem[{Li et~al.(2026)Li, Xu, Ma, Chen, He, Kong, and Chen}]{li2026finelap}
Xiquan Li, Xuenan Xu, Ziyang Ma, Wenxi Chen, Haolin He, Qiuqiang Kong, and Xie Chen. 2026.
\newblock \href {https://arxiv.org/abs/2604.01155} {{FineLAP}: Taming heterogeneous supervision for fine-grained language-audio pretraining}.
\newblock \emph{Preprint}, arXiv:2604.01155.

\bibitem[{Liu et~al.(2025)Liu, Han, Yu, Li, and You}]{liu2025timer1comprehensivetemporalreasoning}
Zijia Liu, Peixuan Han, Haofei Yu, Haoru Li, and Jiaxuan You. 2025.
\newblock \href {https://arxiv.org/abs/2505.13508} {Time-r1: Towards comprehensive temporal reasoning in llms}.
\newblock \emph{Preprint}, arXiv:2505.13508.

\bibitem[{Mei et~al.(2024)Mei, Meng, Liu, Kong, Ko, Zhao, Plumbley, Zou, and Wang}]{mei2024wavcaps}
Xinhao Mei, Chutong Meng, Haohe Liu, Qiuqiang Kong, Tom Ko, Chengqi Zhao, Mark~D Plumbley, Yuexian Zou, and Wenwu Wang. 2024.
\newblock Wavcaps: A chatgpt-assisted weakly-labelled audio captioning dataset for audio-language multimodal research.
\newblock \emph{IEEE/ACM Transactions on Audio, Speech, and Language Processing}, 32:3339--3354.

\bibitem[{Moon et~al.(2023)Moon, Hyun, Park, Park, and Heo}]{moon2023query}
WonJun Moon, Sangeek Hyun, SangUk Park, Dongchan Park, and Jae-Pil Heo. 2023.
\newblock Query-dependent video representation for moment retrieval and highlight detection.
\newblock In \emph{Proceedings of the IEEE/CVF conference on computer vision and pattern recognition}, pages 23023--23033.

\bibitem[{Munakata et~al.(2026)Munakata, Imamura, Nishimura, and Komatsu}]{nakada2025castella}
Hokuto Munakata, Takehiro Imamura, Taichi Nishimura, and Tatsuya Komatsu. 2026.
\newblock {CASTELLA}: Long audio dataset with captions and temporal boundaries.
\newblock In \emph{Proceedings of the IEEE International Conference on Acoustics, Speech and Signal Processing (ICASSP)}.

\bibitem[{Munakata et~al.(2025)Munakata, Nishimura, Nakada, and Komatsu}]{munakata2025amr}
Hokuto Munakata, Taichi Nishimura, Shota Nakada, and Tatsuya Komatsu. 2025.
\newblock Language-based audio moment retrieval.
\newblock In \emph{Proceedings of the IEEE International Conference on Acoustics, Speech and Signal Processing (ICASSP)}.

\bibitem[{Primus et~al.(2025)Primus, Schmid, and Widmer}]{primus2025tacos}
Paul Primus, Florian Schmid, and Gerhard Widmer. 2025.
\newblock \href {https://arxiv.org/abs/2505.07609} {{TACOS}: Temporally-aligned audio captions for language-audio pretraining}.
\newblock \emph{Preprint}, arXiv:2505.07609.

\bibitem[{Radford et~al.(2021)Radford, Kim, Hallacy, Ramesh, Goh, Agarwal, Sastry, Askell, Mishkin, Clark et~al.}]{radford2021learning}
Alec Radford, Jong~Wook Kim, Chris Hallacy, Aditya Ramesh, Gabriel Goh, Sandhini Agarwal, Girish Sastry, Amanda Askell, Pamela Mishkin, Jack Clark, et~al. 2021.
\newblock Learning transferable visual models from natural language supervision.
\newblock \emph{International conference on machine learning}, pages 8748--8763.

\bibitem[{Serizel et~al.(2018)Serizel, Turpault, Eghbalzadeh, and Shah}]{Serizel2018LargeScaleWL}
Romain Serizel, Nicolas Turpault, Hamid Eghbalzadeh, and Ankit Shah. 2018.
\newblock \href {https://api.semanticscholar.org/CorpusID:51864948} {Large-scale weakly labeled semi-supervised sound event detection in domestic environments}.
\newblock \emph{ArXiv}, abs/1807.10501.

\bibitem[{Shao et~al.(2026)Shao, Su, Tian, Mu, Lin, Fan, Luo, Luan, and Xie}]{shao2026listeningwithtime}
Mingchen Shao, Hang Su, Wenjie Tian, Bingshen Mu, Zhennan Lin, Lichun Fan, Zhenbo Luo, Jian Luan, and Lei Xie. 2026.
\newblock \href {https://arxiv.org/abs/2604.22245} {Listening with time: Precise temporal awareness for long-form audio understanding}.
\newblock \emph{Preprint}, arXiv:2604.22245.

\bibitem[{Shao et~al.(2024)Shao, Wang, Zhu, Xu, Song, Bi, Zhang, Zhang, Li, Wu et~al.}]{shao2024deepseekmath}
Zhihong Shao, Peiyi Wang, Qihao Zhu, Runxin Xu, Junxiao Song, Xiao Bi, Haowei Zhang, Mingchuan Zhang, YK~Li, Yang Wu, et~al. 2024.
\newblock Deepseekmath: Pushing the limits of mathematical reasoning in open language models.
\newblock \emph{arXiv preprint arXiv:2402.03300}.

\bibitem[{Sun et~al.(2026)Sun, Zhou, Li, Zhang, Wang, and Xie}]{sun2026spotsoundenhancinglargeaudiolanguage}
Luoyi Sun, Xiao Zhou, Zeqian Li, Ya~Zhang, Yanfeng Wang, and Weidi Xie. 2026.
\newblock \href {https://arxiv.org/abs/2604.13023} {Spotsound: Enhancing large audio-language models with fine-grained temporal grounding}.
\newblock \emph{Preprint}, arXiv:2604.13023.

\bibitem[{Tang et~al.(2024)Tang, Yu, Sun, Chen, Tan, Li, Lu, MA, and Zhang}]{tang2024salmonn}
Changli Tang, Wenyi Yu, Guangzhi Sun, Xianzhao Chen, Tian Tan, Wei Li, Lu~Lu, Zejun MA, and Chao Zhang. 2024.
\newblock \href {https://openreview.net/forum?id=14rn7HpKVk} {Salmonn: Towards generic hearing abilities for large language models}.
\newblock In \emph{The Twelfth International Conference on Learning Representations}.

\bibitem[{Turpault et~al.(2019)Turpault, Serizel, Shah, and Salamon}]{turpault2019sound}
Nicolas Turpault, Romain Serizel, Ankit~Parag Shah, and Justin Salamon. 2019.
\newblock Sound event detection in domestic environments with weakly labeled data and soundscape synthesis.
\newblock In \emph{Workshop on Detection and Classification of Acoustic Scenes and Events}.

\bibitem[{Vyas et~al.(2025)Vyas, Chang, Yang, Huang, Gao, Richter, Chen, Le, Dollár, Feichtenhofer, Lee, and Hsu}]{vyas2025pushingfrontieraudiovisualperception}
Apoorv Vyas, Heng-Jui Chang, Cheng-Fu Yang, Po-Yao Huang, Luya Gao, Julius Richter, Sanyuan Chen, Matt Le, Piotr Dollár, Christoph Feichtenhofer, Ann Lee, and Wei-Ning Hsu. 2025.
\newblock \href {https://arxiv.org/abs/2512.19687} {Pushing the frontier of audiovisual perception with large-scale multimodal correspondence learning}.
\newblock \emph{Preprint}, arXiv:2512.19687.

\bibitem[{Wang et~al.(2025)Wang, Li, Ma, Liu, and Wang}]{wang2025timeaudio}
Hualei Wang, Yiming Li, Shuo Ma, Hong Liu, and Xiangdong Wang. 2025.
\newblock Listening between the frames: Bridging temporal gaps in large audio-language models.
\newblock \emph{arXiv preprint arXiv:2511.11039}.

\bibitem[{Wu et~al.(2024)Wu, Chen, Zhang, Hui, Nezhurina, Berg-Kirkpatrick, and Dubnov}]{wu2024largescalecontrastivelanguageaudiopretraining}
Yusong Wu, Ke~Chen, Tianyu Zhang, Yuchen Hui, Marianna Nezhurina, Taylor Berg-Kirkpatrick, and Shlomo Dubnov. 2024.
\newblock \href {https://arxiv.org/abs/2211.06687} {Large-scale contrastive language-audio pretraining with feature fusion and keyword-to-caption augmentation}.
\newblock \emph{Preprint}, arXiv:2211.06687.

\bibitem[{Xie et~al.(2024{\natexlab{a}})Xie, Xu, Wu, and Wu}]{xie2024audiotime}
Zeyu Xie, Xuenan Xu, Zhizheng Wu, and Mengyue Wu. 2024{\natexlab{a}}.
\newblock \href {https://arxiv.org/abs/2407.02857} {{AudioTime}: A temporally-aligned audio-text benchmark dataset}.
\newblock \emph{Preprint}, arXiv:2407.02857.

\bibitem[{Xie et~al.(2024{\natexlab{b}})Xie, Xu, Wu, and Wu}]{xie2024picoaudio}
Zeyu Xie, Xuenan Xu, Zhizheng Wu, and Mengyue Wu. 2024{\natexlab{b}}.
\newblock \href {https://arxiv.org/abs/2407.02869} {{PicoAudio}: Enabling precise timestamp and frequency controllability of audio events in text-to-audio generation}.
\newblock \emph{Preprint}, arXiv:2407.02869.

\bibitem[{Xu et~al.(2025)Xu, Guo, Hu, Chu, Wang, He, Wang, Shi, He, Zhu, Lv, Wang, Guo, Wang, Ma, Zhang, Zhang, Hao, Guo, Yang, Zhang, Ma, Wei, Bai, Chen, Liu, Wang, Yang, Liu, Ren, Zheng, Men, Zhou, Yu, Yang, Yu, Zhou, and Lin}]{xu2025qwen3omnitechnicalreport}
Jin Xu, Zhifang Guo, Hangrui Hu, Yunfei Chu, Xiong Wang, Jinzheng He, Yuxuan Wang, Xian Shi, Ting He, Xinfa Zhu, Yuanjun Lv, Yongqi Wang, Dake Guo, He~Wang, Linhan Ma, Pei Zhang, Xinyu Zhang, Hongkun Hao, Zishan Guo, and 19 others. 2025.
\newblock \href {https://arxiv.org/abs/2509.17765} {Qwen3-omni technical report}.
\newblock \emph{Preprint}, arXiv:2509.17765.

\bibitem[{Xu et~al.(2021)Xu, Dinkel, Wu, and Yu}]{xu2021tag}
Xuenan Xu, Heinrich Dinkel, Mengyue Wu, and Kai Yu. 2021.
\newblock Text-to-audio grounding: Building correspondence between captions and sound events.
\newblock In \emph{Proceedings of the IEEE International Conference on Acoustics, Speech and Signal Processing (ICASSP)}, pages 606--610.

\bibitem[{Xu et~al.(2024)Xu, Ma, Wu, and Yu}]{xu2024wstag}
Xuenan Xu, Ziyang Ma, Mengyue Wu, and Kai Yu. 2024.
\newblock Towards weakly supervised text-to-audio grounding.
\newblock \emph{IEEE Transactions on Multimedia}.

\bibitem[{Yuan et~al.(2019)Yuan, Ma, Wang, Liu, and Zhu}]{yuan2019semantic}
Yitian Yuan, Lin Ma, Jingwen Wang, Wei Liu, and Wenwu Zhu. 2019.
\newblock Semantic conditioned dynamic modulation for temporal sentence grounding in videos.
\newblock \emph{Advances in Neural Information Processing Systems}, 32.

\bibitem[{Zhang et~al.(2020)Zhang, Sun, Jing, and Zhou}]{zhang2020span}
Hao Zhang, Aixin Sun, Wei Jing, and Joey~Tianyi Zhou. 2020.
\newblock Span-based localizing network for natural language video localization.
\newblock In \emph{Proceedings of the 58th annual meeting of the association for computational linguistics}, pages 6543--6554.

\end{thebibliography}
\clearpage
\appendix
\raggedbottom
% Appendix-only inline floats keep each compact result beside its discussion.
\makeatletter
\newenvironment{AppendixTable}{\par\addvspace{8pt}\noindent\begin{minipage}{\columnwidth}\def\@captype{table}}{\end{minipage}\par\addvspace{8pt}}
\newenvironment{AppendixFigure}{\par\addvspace{8pt}\noindent\begin{minipage}{\columnwidth}\def\@captype{figure}}{\end{minipage}\par\addvspace{8pt}}
\makeatother

\section{Stage 1 Synthesis Details}
\label{app:synth}

\paragraph{Pool source.}
The Stage~1 pool comprises ${\sim}122$k real clips from
FSD50K~\citep{fonseca2022fsd50k} and AudioSet~\citep{45857}
% , and a small YouTube set
, disjoint from the FreeSound clips used in Stage~2. For every
pool clip we precompute an active span by per-frame RMS-energy
localization in dBFS (the onset and offset of the loudest sustained region under
the clip's tag); each pool entry thus stores its audio, tag label, and an
RMS-localized active span. Both target and distractor audio are drawn from these
active spans, so every constituent of a synthetic clip is a real recording while
every placement is fixed by the synthesizer.

\paragraph{Composition.}
Each synthetic clip is constructed as follows:
\begin{enumerate}[label=(\arabic*)]
  \item randomly select a target label from the pool;
  \item cut 1--5 copies of that label's active span and place them at
        non-overlapping onsets (inter-segment gap uniformly drawn from
        $[0.5, 4.0]$\,s);
  \item optionally insert 0--2 distractor events of a different label,
        drawn from the same pool and allowed to overlap freely;
  \item mix onto a Gaussian-noise background with foreground SNR uniformly
        drawn from $[10, 20]$\,dB;
  \item set total clip duration to $[10, 30]$\,s at 16\,kHz.
\end{enumerate}
The occurrence-count distribution is deliberately skewed toward multiple events:
1 (20\%), 2 (30\%), 3 (25\%), 4 (15\%), 5 (10\%). The distractor-count
distribution is 0 (30\%), 1 (50\%), 2 (20\%). Clips are stored at 16\,kHz in FLAC
format, and the ground-truth segment list is the exact set of placement
onsets/offsets the synthesizer wrote, rounded to three decimals.

Each clip is paired with a List-All query:
\begin{quote}\small
\textit{``The audio contains the sound event: \texttt{[label]}. List ALL time
intervals when this event occurs. Format your answer as a JSON array of
[start, end] pairs in seconds.''}
\end{quote}
with response \texttt{<answer>[[s1,e1],[s2,e2],...]\allowbreak</answer>}.
A short \texttt{<think>} prefix is prepended during training to encourage
step-by-step temporal reasoning.

\section{Stage 2 Annotation Details}
\label{app:annot}

The Stage~2 training-data pipeline processes
10{,}000 candidate FreeSound clips before filtering; this candidate pool is distinct from the retained GRPO corpus.
Audio labeling uses \texttt{gemini-3-pro-preview}; all text-only judgements use
\texttt{gemini-2.5-flash}; temporal localization uses PE~A-Frame
(\texttt{pe-a-frame-large}).

\paragraph{Label identification.}
The audio is chunked into 10\,s segments (chunking mitigates the tendency of
LALMs to miss events in long recordings), and each chunk is sent to
\texttt{gemini-3-pro-preview}, which returns the clearly audible events in JSON
(Figure~\ref{fig:prompt_label}). The prompt requests at most eight
semantically orthogonal labels (1--3 words, lowercase), prefers the most specific
descriptor over a parent category (\textit{dog barking} over \textit{animal
sound}), and abstains from generic ambient labels (\textit{background noise},
\textit{music}, \textit{wind}) unless such a sound is itself the sole dominant
event. Chunk-level labels are pooled and lightly canonicalized (character
normalization, deduplication, stop-list removal) into a per-clip inventory.

\paragraph{Event-type classification.}
Each label is classified as continuous (sustained events such as engine
noise or rain) or discrete (transient events such as a knock or a bark)
via a text-only \texttt{gemini-2.5-flash} query (see Figure~\ref{fig:prompt_clean}),
cached per unique label string. This tag controls whether nearby PE~A-Frame
spans are merged: a discrete event tagged continuous may have separate
occurrences merged, while a continuous event tagged discrete may be fragmented.

\paragraph{Temporal localization.}
PE A-Frame computes per-frame audio-text similarity at $\approx$40\,ms
resolution over the full clip; frames whose score exceeds $0.5$ are marked active
and consecutive active frames are merged into onset/offset spans. For continuous
labels, adjacent spans separated by at most $0.5$\,s are further merged to
suppress brief dropouts; for discrete labels no merging is applied, so each peak
yields a distinct occurrence. Labels for which no active frame is found are
discarded, ensuring every retained span reflects a confident detection.

\paragraph{Global label cleaning.}
Because labels are coined independently per clip, near-synonyms
(\textit{car engine}, \textit{engine noise}, \textit{vehicle engine}) may
coexist. We collect all unique labels, embed each with CLAP~\citep{wu2024largescalecontrastivelanguageaudiopretraining}, 
and cluster the embeddings by
$k$-means with $k=\lceil N/10\rceil$ (on average $\leq$10 labels per cluster).
Each cluster, with per-label occurrence counts, is sent to Gemini, which assigns
every label keep (specific and non-redundant), drop (generic or
duplicating a more specific label in the same cluster), or rename
(replace by a provided canonical form; Figure~\ref{fig:prompt_clean}, bottom).
The resulting global mapping is applied to every record.

\paragraph{Retention and query expansion.}
After Phase~1 and Phase~2, 5{,}742 of the 10{,}000 annotated clips carry at least
one Gemini category and one localized PE~A-Frame span. We retain 2{,}000
multi-event clips containing at least two distinct event categories; a generic tag-confidence score does not discriminate
among candidates (mean $0.96$, near-saturated) and is not used for ranking. Each
retained clip contributes one List-All query per category, expanding the 2{,}000
clips to 5{,}244 positive (clip, category) corpus queries. The training loader retains target intervals of at least 0.2 seconds and drops queries with no remaining target interval. This removes 334 queries, leaving 4{,}910 GRPO training queries from 1{,}987 recordings, consistent with the per-epoch training logs. The consolidated corpus statistics in Table~\ref{tab:data_stats} report the training queries and recordings after this filtering.

\begin{figure*}[t]
\centering
\begin{minipage}{0.98\textwidth}
\small
\rule{\linewidth}{0.6pt}
\begin{verbatim}
Identify clearly recognizable sound events in this audio clip.
Reply ONLY with JSON: {"sounds": ["label1", "label2", ...]}.
Return at most 8 labels.
Important rules:
(1) Labels should be semantically orthogonal: do not output multiple labels that
    point to the same event family. For example, choose only one of
    "bird chirping", "bird singing", "bird vocalization"; choose only one of
    "bus", "motor vehicle", "vehicle".
(2) Prefer the most specific audible event label, not its parent category.
(3) Do NOT output generic ambience-only labels such as background noise, noise,
    human voice, sound effect, music, water, or wind, unless that generic sound
    is itself the single dominant event.
(4) If unsure whether a label is distinct or dominant enough, abstain. Prefer
    fewer labels.
(5) Use short common English phrases, lowercase style.
Reply with compact minified JSON only. Do not use markdown, code fences,
explanations, or extra keys.
\end{verbatim}
\rule{\linewidth}{0.6pt}
\end{minipage}
\caption{Label-identification prompt, sent once per 10\,s audio chunk; labels
  from all chunks are unioned into the per-clip inventory.}
\label{fig:prompt_label}
\end{figure*}

\begin{figure*}[t]
\centering
\begin{minipage}{0.98\textwidth}
\small
\rule{\linewidth}{0.6pt}

\textbf{Event-type classification (per unique label, text-only).}
\begin{verbatim}
Is the sound "{label}" continuous (sustained over time, e.g. engine noise, rain,
music, wind) or discrete (short events, e.g. knock, clap, beep, gunshot)?
Reply with only one word: continuous or discrete.
\end{verbatim}

\textbf{Cluster-level label cleaning (per $k$-means cluster).} Here
\texttt{\{items\}} is a comma-separated list of \texttt{"label" (n=count)} pairs.
\begin{verbatim}
You are an audio taxonomy expert. Here is a cluster of sound labels: [{items}].
For each label decide: keep (it is specific and useful), drop (it is
generic/redundant), or rename (provide a better canonical form).
Reply ONLY with JSON: {"decisions": {
    "label": {"action": "keep"|"drop"|"rename", "rename_to": "new_name_or_null"}
}}.
Lowercase phrases.
\end{verbatim}
\rule{\linewidth}{0.6pt}
\end{minipage}
\caption{Event-type classification (top) and global label-cleaning (bottom)
  prompts. The classification result controls whether adjacent PE A-Frame spans
  are merged (continuous) or kept separate (discrete).}\label{fig:prompt_clean}
\end{figure*}

\section{Independent Annotation and Adjudication}
\label{app:agreement}

Three non-author volunteers independently label every candidate audio clip from scratch,
without seeing one another's annotations. Using a shared written protocol,
each identifies the audible sound categories and marks all occurrences as
$[t_{\mathrm{start}},t_{\mathrm{end}}]$ pairs in seconds. The initial category
inventory is the AudioSet Ontology; labels for sounds outside that inventory
are adopted by majority vote among the three annotators. The volunteers gave informed consent to participate. Under the institution's
policy, this type of annotation did not require ethics-board review.

After independent annotation, corresponding occurrences are compared using
interval IoU. When the proposed boundaries agree at IoU $>0.95$, their onset
and offset times are averaged. Disagreements at IoU $\leq0.95$ are referred to a
senior annotator, who listens again and determines the final interval. Cases
that remain unresolved after adjudication (2.1\%) are excluded. This procedure
establishes the evaluation references directly from human listening and
boundary annotation.

\section{AEGBench Construction Details}
\label{app:bench}

\paragraph{Sources and annotation.}
AEGBench combines audio from AudioSet Strong Labels, FSD50K eval~\citep{fonseca2022fsd50k}, the BBC Sound Effects
Library, and YouTube Life Sounds. Clip durations range from 10 to 120\,s.
Every clip is annotated from scratch by three annotators working independently.
The AudioSet Ontology supplies the initial label inventory. When an audible
sound falls outside this inventory, the annotators decide its new label by
majority vote. Appendix~\ref{app:agreement} describes boundary reconciliation.

\paragraph{Selection criteria.}
Candidates satisfy energy contrast at least 12\,dB, active ratio in
$[0.10,0.85]$, and target-event duration in $[0.5,60]$\,s, with at most
40 items per fine-grained category. Difficulty results average the queries selected by the target-conditioned or recording-context rules below, without additional per-tag subsampling. Tags are non-exclusive.

\paragraph{Difficulty-subset tags.}
PO, RO, and LD are conditioned on the recording--target pair, using the independently annotated reference intervals:
\begin{enumerate}
  \item \textbf{Polyphonic Overlap (PO)}: in a released PO-tagged recording, at least one interval of the queried target overlaps an interval of another category by more than $0.05$\,s.
  \item \textbf{Repeated Occurrence (RO)}: the queried target has at least three distinct annotated occurrences, matching the released recording-level RO criterion.
  \item \textbf{Long Duration (LD)}: at least one interval of the queried target lasts more than $30$\,s.
\end{enumerate}
An unrelated category cannot place a query in these target-conditioned subsets. GO, NS, and SA retain their released recording-level labels for gradual transitions, low-contrast activity, and semantic ambiguity. All queries in such a recording enter the corresponding context subset, so these results concern the recording environment rather than a verified property of each target.

\paragraph{Source and subset counts.}
Table~\ref{tab:bench_source_counts} reports the released query-row distribution. Exact duplicate reference intervals are removed within a query, while released query-row weights are retained. Subsets are non-exclusive.

\begin{AppendixTable}
\centering\footnotesize
\setlength{\tabcolsep}{3pt}
\begin{tabular}{@{}lrrr@{}}
\toprule
Source & Recordings & Queries & Events \\
\midrule
AudioSet Strong & 2,228 & 6,819 & 11,121 \\
FSD50K eval & 954 & 2,276 & 4,190 \\
BBC Sound Effects & 234 & 715 & 2,396 \\
YouTube Life Sounds & 8 & 114 & 892 \\
\midrule
Total & 3,424 & 9,924 & 18,599 \\
\bottomrule
\end{tabular}
\caption{AEGBench source distribution. Recordings are unique audio IDs; queries and events retain the released row weights. The manifest contains 3,425 recording entries and 9,911 unique recording--target pairs.}
\label{tab:bench_source_counts}
\end{AppendixTable}

\begin{AppendixTable}
\centering\footnotesize
\setlength{\tabcolsep}{4pt}
\begin{tabular}{@{}llrr@{}}
\toprule
Subset & Unit & Recordings & Queries \\
\midrule
PO & Target & 2,295 & 8,193 \\
RO & Target & 1,277 & 1,738 \\
LD & Target & 53 & 80 \\
GO & Recording & 2,472 & 7,039 \\
NS & Recording & 1,617 & 4,570 \\
SA & Recording & 1,233 & 5,636 \\
\bottomrule
\end{tabular}
\caption{Difficulty-subset membership. Recordings count unique audio IDs; queries retain the released row weights. A query can belong to multiple subsets.}
\label{tab:bench_subset_counts}
\end{AppendixTable}

\paragraph{Comparison with existing resources.}
Compared with phrase-level AudioGrounding~\citep{xu2021tag} and AMR~\citep{munakata2025amr}, AEGBench emphasizes multi-occurrence prediction and difficulty-specific diagnostics. CASTELLA~\citep{nakada2025castella} and long-form audio benchmarks~\citep{shao2026listeningwithtime} extend temporal coverage, while TACOS~\citep{primus2025tacos} supplies a larger grounding corpus. AEGBench complements these resources through independently annotated timestamps, an extensible event vocabulary across diverse sources, and explicit hard-case subsets. The Auto-AEG training annotations form a separate resource.
Table~\ref{tab:benchmark_comparison} summarizes the complementary evaluation emphases of these resources.

\begin{AppendixTable}
\centering
\small
\setlength{\tabcolsep}{3pt}
\begin{tabularx}{\linewidth}{@{}p{0.30\linewidth}X@{}}
\toprule
\textbf{Resource} & \textbf{Evaluation emphasis} \\
\midrule
DCASE Task 4 & Strong-label sound event detection over a fixed class inventory. \\
AudioGrounding & Phrase-conditioned temporal localization. \\
AMR & Language-based moment retrieval with simulated audio. \\
CASTELLA & Temporal grounding in long-form real recordings. \\
TACOS & Large-scale text--audio grounding with human temporal annotations. \\
\textbf{AEGBench} & \textbf{All-occurrence event grounding with explicit difficulty subsets.} \\
\bottomrule
\end{tabularx}
\caption{Comparison with existing temporal audio resources. AEGBench's evaluation annotations and the Auto-AEG training corpus are separate resources.}
\label{tab:benchmark_comparison}
\end{AppendixTable}

\section{Training Configuration}
\label{app:train_config}

Both stages fine-tune Q3-Omni with QLoRA: each backbone is loaded
in 4-bit NF4 with bfloat16 compute dtype, and LoRA adapters (rank $r{=}16$,
$\alpha{=}32$) target the
$\{q,k,v,o,\text{gate},\text{up},\text{down}\}\_\text{proj}$ modules of the
language-model component. The audio tower is frozen and kept in bfloat16. All
runs use a single GPU. Stage-specific hyperparameters are listed in
Table~\ref{tab:train_config}.

The main SFT and GRPO stages require approximately 120 A100-equivalent
GPU-hours in a comparable-run extrapolation. The original v1 runs used one
80\,GB GPU, but their archived records do not preserve wall-clock duration or
the exact GPU model. To estimate compute, we scale observed A100-SXM4-80GB
training time from Q3-Omni reference runs by the v1 training-set sizes: a
10{,}800-item, three-epoch SFT run used 2 GPUs for 10.0 hours, and a
1{,}024-item, three-epoch GRPO run used 4 GPUs for 5.14 hours. Scaling to
9{,}000 SFT items and 4{,}910 GRPO items gives approximately 17 and 99
GPU-hours, respectively. This estimate covers the main training stages;
evaluation and ablation runs are excluded.

The continued-SFT control uses the same synthetic initialization and real-data query pool. Its training configuration uses learning rate $2\times10^{-4}$, three epochs, and a 90:10 training/validation split. The reported control evaluates the third-epoch checkpoint rather than the separately saved validation-loss-best adapter.

\begin{AppendixTable}
  \centering
  \footnotesize
  \setlength{\tabcolsep}{3pt}
  
  \begin{tabular}{@{}p{0.28\columnwidth}p{0.29\columnwidth}p{0.35\columnwidth}@{}}
    \toprule
     & \textbf{Stage 1: SFT} & \textbf{Stage 2: GRPO} \\
    \midrule
    Training data             & 10{,}000 synth.\ clips  & 4{,}910 real queries \\
    Train/val split           & 9:1                     & all \\
    Epochs                    & 3                        & 3 \\
    Learning rate             & $2{\times}10^{-4}$       & $5{\times}10^{-5}$ \\
    Batch / effective         & 2 / 16 (grad.\ accum.)   & 1 prompt $\times$ 4 rollouts \\
    Rollouts per sample ($G$) & N/A                     & 4 \\
    Regularizer               & weight decay, $1{\times}10^{-2}$                     & L2 to SFT adapter, $\beta{=}0.04$ \\
    Checkpoint                & best by val.\ loss       & final \\
    \bottomrule
  \end{tabular}
  
  \caption{Stage-specific training hyperparameters. Both stages share the
    single-GPU QLoRA-NF4 / LoRA ($r{=}16$, $\alpha{=}32$) configuration
    described above.}
  \label{tab:train_config}
\end{AppendixTable}

\section{Reward Implementation}
\label{app:reward_settings}

The GRPO reward combines four weighted terms,
$r = w_{\text{iou}}\,r_{\text{iou}} + w_{\text{fmt}}\,r_{\text{fmt}}
+ w_{\text{nem}}\,r_{\text{nem}} + w_{\text{prec}}\,r_{\text{prec}}$
(see the Reward Function paragraph of the main paper).
The default weights are $(w_{\text{iou}},w_{\text{fmt}},w_{\text{nem}},w_{\text{prec}})
=(0.65,0.15,0.05,0.15)$. The format term awards $0.3$ for each required
\texttt{<think>} and \texttt{<answer>} block and $0.4$ for a valid JSON array.
The non-empty term equals $1$ for at least one predicted interval. The
count-based precision term decreases linearly once the predicted-to-reference
count ratio exceeds $2$, reaching zero at a ratio of $4$.
Table~\ref{tab:reward_components} in the main paper removes each reward term in turn while retaining the other weights.

\section{Audio-Length Handling in the AEGBench Evaluation}
\label{app:audio_length}

AEGBench clips span 10--120\,s. The evaluation loader reads the entire waveform,
converts it to mono at 16\,kHz, and forwards it without cropping. All metrics use
the full annotated duration, so events missed near the end of a recording
remain in the reference set.

Qwen3-Omni encodes audio at approximately 26 tokens/s and chunks features with
an 800-frame attention window; its token count scales with recording length.
The external LALMs receive the complete waveform through their native
front-ends: Qwen2-Audio's Whisper-family encoder, Audio Flamingo 3's
audio encoder, Kimi-Audio's encoder, and Gemini-3-Pro's server-side processing.
DASM performs frame-level scoring over the recording using the overlapping
window configuration in Appendix~\ref{app:detectors}. Any internal model
length restriction therefore belongs to the evaluated system; the benchmark
loader imposes no 30\,s crop.

\section{Evaluation Metrics}
\label{app:metrics}

Denote ground-truth and predicted intervals by
$\mathcal{Y}^*=\{[s^*_j,e^*_j]\}_{j=1}^{M}$ and
$\hat{\mathcal{Y}}=\{[s_i,e_i]\}_{i=1}^{K}$.

\paragraph{mIoU.}
The mean IoU between each ground-truth interval and its best-matching
prediction, averaged over evaluation queries ($|\mathcal{Y}^*|\geq 1$):
\begin{equation}
  \text{mIoU} = \frac{1}{|\mathcal{Y}^*|}\sum_{j}
    \max_{i}\operatorname{IoU}\!\left([s^*_j,e^*_j],\,[s_i,e_i]\right).
\end{equation}
This is a recall-oriented localization score: each ground-truth segment
is credited for the prediction that best overlaps it, and extra or hallucinated
intervals are not penalized. To control for over-prediction we additionally
report Interval F1@0.5 and interval precision, which use one-to-one matching
and explicitly penalize spurious predictions; the AEGBench improvements hold
under all three.

\paragraph{Interval IoU.}
For two intervals $A=[a_s,a_e]$ and $B=[b_s,b_e]$,
$\operatorname{IoU}(A,B)=|A\cap B|/|A\cup B|$, with
$|A\cap B|=\max(0,\min(a_e,b_e)-\max(a_s,b_s))$ and
$|A\cup B|=(a_e-a_s)+(b_e-b_s)-|A\cap B|$.

\paragraph{Recall-/Precision-/F1-IoU@$\theta$.}
With $M{=}|\mathcal{Y}^*|$ ground-truth and $K{=}|\hat{\mathcal{Y}}|$ predicted
intervals,
\begin{equation}
  \begin{split}
\text{R@}\theta=\tfrac{1}{M}\textstyle\sum_{j}\mathbf{1}\!\left[\max_i\operatorname{IoU}(Y^*_j,\hat Y_i)\ge\theta\right],\\
\text{P@}\theta=\tfrac{1}{K}\textstyle\sum_{i}\mathbf{1}\!\left[\max_j\operatorname{IoU}(\hat Y_i,Y^*_j)\ge\theta\right],
  \end{split}
\end{equation}
Their harmonic mean, F1-IoU@$\theta$, is the localization component of the GRPO reward.
The soft analog, mIoU, averages the best-match IoU
$\max_i\operatorname{IoU}(Y^*_j,\hat Y_i)$ rather than thresholding it, so
$\text{mIoU}\ge\theta\,\text{R@}\theta$ at the same $\theta$.

\paragraph{Onset F1 and onset P/R.}
Additional archived onset diagnostics use matching with collar
$\delta=0.5$\,s. The scorer visits predicted onsets in their output order
and greedily matches each to the nearest still-unmatched reference onset
within the collar. Offsets do not affect this metric. With the resulting
one-to-one TP count, onset precision is $\mathrm{TP}/K$, recall is
$\mathrm{TP}/M$, and Onset F1 is $2\mathrm{TP}/(K+M)$.
The archived \texttt{ev\_F1} field denotes this Onset F1, not interval
matching. For a positive query with no predictions, the legacy scorer
sets precision to one and recall and F1 to zero.

\paragraph{Interval F1@0.5 (IF1).}
All primary benchmark comparisons and the 7B comparison use one-to-one interval matching. A Hungarian
assignment maximizes the number of pairs with $\operatorname{IoU}\geq0.5$.
For each query, $\mathrm{FP}=K-\mathrm{TP}$ and
$\mathrm{FN}=M-\mathrm{TP}$. The reported micro precision and F1 are
\begin{equation}
\begin{aligned}
 P_{\rm micro}&=\frac{\sum_q\mathrm{TP}_q}
 {\sum_q(\mathrm{TP}_q+\mathrm{FP}_q)},\\
 F1_{\rm micro}&=\frac{2\sum_q\mathrm{TP}_q}
 {2\sum_q\mathrm{TP}_q+\sum_q\mathrm{FP}_q+\sum_q\mathrm{FN}_q}.
\end{aligned}
\end{equation}
This evaluation criterion differs from the many-to-one F1-IoU component
used in the GRPO reward.

\paragraph{Segment F1.}
The scorer discretizes the timeline into a one-second grid and marks bins
covered by at least one interval as active. Segment F1 compares the unions
of predicted and reference activity, using the resulting bin-level TP,
FP, and FN. It is calculated separately for each query and then averaged;
it is not pooled across recordings.

\paragraph{Aggregation.}
mIoU first averages best-match IoU over the reference intervals within a query, then averages equally over queries. R@$\theta$ and one-second Segment F1 also use query-macro means. Primary Interval F1/P/R use micro aggregation of the one-to-one TP/FP/FN counts; no class weighting is applied. The released AEGBench inventory contains 9,924 queries and 18,599 reference events after exact within-query duplicate removal. AudioGrounding uses the complete 997 queries and 1,822 reference events. Evaluation uses full recording durations. Each query's exact duplicate reference intervals are removed, without merging neighboring events or clipping predicted endpoints. Generation failures count as empty predictions. Every AEGBench main-table model and the continued-SFT control cover all 9,924 queries with the same 18,599 reference events. Additional onset diagnostics retain their archived query-macro aggregation. The reward-component ablation retains its archived mIoU/R@0.5 and onset-precision diagnostics; these are distinct from interval F1.

\section{Recovery under Different IoU Criteria}
\label{app:riou_sweep}

Table~\ref{tab:riou_sweep} rescores the same fixed predictions at IoU criteria $0.3$ and $0.7$, complementing R@0.5 in the main results. Q3-Omni improves through both adaptation stages and retains the highest recovery under both criteria. PE-A-Frame is competitive at the stricter criterion, whereas TEMPO's GRPO checkpoint improves on its SFT checkpoint throughout. These results distinguish event coverage from the one-to-one interval matching reported in the primary tables.

\begin{AppendixTable}
\centering\small
\begin{tabular}{@{}lrr@{}}
\toprule
Model & R@0.3 & R@0.7 \\
\midrule
Gemini-3-Pro & 0.464 & 0.154 \\
Kimi-Audio-7B & 0.157 & 0.018 \\
Qwen2-Audio-7B & 0.226 & 0.025 \\
Audio Flamingo 3 & 0.329 & 0.070 \\
CLAP & 0.388 & 0.152 \\
TAG & 0.314 & 0.134 \\
DASM & 0.260 & 0.164 \\
PE-A-Frame & 0.470 & 0.304 \\
TEMPO SFT & 0.506 & 0.242 \\
TEMPO + GRPO & 0.571 & 0.272 \\
Q3 base & 0.397 & 0.127 \\
Q3 + SFT & 0.519 & 0.177 \\
Q3 + GRPO & \textbf{0.624} & \textbf{0.351} \\
\bottomrule
\end{tabular}
\caption{Query-macro Recall-IoU under two criteria, using fixed predictions on all 9,924 AEGBench queries.}
\label{tab:riou_sweep}
\end{AppendixTable}

\section{Additional Onset Metrics for the Training Ablation}
\label{app:ablation_onset}

Table~\ref{tab:ablation_onset} retains the archived onset precision and recall diagnostics for the continued-SFT control in Section~\ref{sec:ablation}. All settings use the same
Q3-Omni backbone and synthetic cold-start; the two Stage~2 variants use the
same retained real-data corpus.

\begin{AppendixTable}
  \centering
  \small
  \begin{tabular}{@{}lcc@{}}
    \toprule
    \textbf{Setting} & \textbf{OP} & \textbf{OR} \\
    \midrule
    Stage 1 SFT & 0.368 & 0.387 \\
    + Stage 2 SFT & 0.386 & 0.425 \\
    + Stage 2 GRPO & \textbf{0.559} & \textbf{0.566} \\
    \bottomrule
  \end{tabular}
  \caption{Archived query-macro onset diagnostics for the training-objective comparison on AEGBench. OP/OR denote onset precision/recall with a $0.5$\,s collar.}
  \label{tab:ablation_onset}
\end{AppendixTable}

\section{Training Prompt Template}
\label{app:train_prompt}

Figure~\ref{fig:train_prompt} gives the shared prompt and the supervised response scaffold.

\begin{figure*}[t]
\centering
\begin{minipage}{0.98\textwidth}
\small
\rule{\linewidth}{0.6pt}

\textbf{System prompt (shared by SFT and GRPO).}
\begin{verbatim}
You are an audio temporal grounding assistant. Listen carefully to the audio
and answer questions about when specific sound events occur. Think step by step
inside <think> tags, then give your final answer inside <answer> tags.
\end{verbatim}

\textbf{User query (List-All).}
\begin{verbatim}
The audio contains the sound event: "{category}". List ALL time intervals
(start, end in seconds) when this event occurs. Reply ONLY with a JSON array
inside <answer> tags, e.g. <answer>[[0.5, 2.1], [5.0, 7.3]]</answer> or
<answer>[]</answer> if not present.
\end{verbatim}

\textbf{Gold response with reasoning (SFT target).}
\begin{verbatim}
<think>
I need to find all time intervals when {category} occurs in this audio
(total duration: {duration}s). Listening carefully, I can identify
{n} occurrences: {parts}.
</think>
<answer>{ans}</answer>
\end{verbatim}
\rule{\linewidth}{0.6pt}
\end{minipage}
\caption{Prompt template shared by SFT and GRPO. The system prompt elicits a
  \texttt{<think>} reasoning trace before the \texttt{<answer>}. During SFT the
  model is trained on gold responses whose chain-of-thought enumerates every
  occurrence (the single-occurrence target is identical with ``one occurrence'');
  during GRPO the same system prompt and query are used, but the response is
  sampled and scored by the interval-aware reward rather than supervised. A
  no-rejection query variant (omitting the \texttt{<answer>[]} option) is used
  for clips where the target event is guaranteed present.}
\label{fig:train_prompt}
\end{figure*}

\section{Reward Transfer}
\label{app:reward_7b}

We repeat the SFT-to-GRPO experiment with Qwen2.5-Omni-7B. The two GRPO
arms use the same SFT initialization, positive real-audio queries, and
optimization settings; the alternative sets only the auxiliary
count-based precision weight $w_{\text{prec}}$ from $0.15$ to $0$, without
renormalizing the remaining weights. The F1-IoU localization term still
penalizes unmatched predictions. We evaluate the unadapted base model, the SFT checkpoint, and both
GRPO arms with the same prompts, decoding, and interval scorer. Both datasets use their complete evaluation query sets; failed generations are scored as empty predictions.

\begin{AppendixTable}
  \centering\footnotesize
  \setlength{\tabcolsep}{2pt}
  \begin{tabular}{@{}lrrrr@{}}
    \toprule
    & \multicolumn{2}{c}{\textbf{DESED}} & \multicolumn{2}{c}{\textbf{AEGBench}} \\
    \cmidrule(lr){2-3}\cmidrule(l){4-5}
    \textbf{7B checkpoint} & \textbf{mIoU} & \textbf{IF1} & \textbf{mIoU} & \textbf{IF1} \\
    \midrule
    Base & 0.497 & 0.330 & 0.331 & 0.138 \\
    SFT & 0.520 & 0.328  & 0.440 & 0.268 \\
    Standard GRPO, epoch 2 & 0.485 & 0.299  & 0.444 & 0.334 \\
    Standard GRPO, epoch 3 & 0.422 & 0.321  & 0.348 & 0.238 \\
    $w_{\text{prec}}=0$, epoch 3 & \textbf{0.573} & \textbf{0.454}  & \textbf{0.461} & \textbf{0.357} \\
    \bottomrule
  \end{tabular}
  \caption{Base, SFT, and reward comparison on Qwen2.5-Omni-7B. Interval F1@0.5 uses one-to-one
  matching at IoU $\geq0.5$ and micro aggregation. All rows use the same 1,112 DESED queries or
  the complete 9,924-query AEGBench evaluation set.}
  \label{tab:reward_7b}
\end{AppendixTable}

SFT improves temporal overlap over the unadapted base on both datasets
and substantially improves strict event matching on AEGBench, while
DESED event F1 remains largely unchanged. Removing the extra precision
term yields further gains over SFT in both overlap and strict event
matching across the two evaluations. The standard-reward run weakens in its
third epoch, coinciding with a sharp rise in its training KL; its second
epoch also trails the zero-precision arm. Thus, the additional precision
bonus is not uniformly beneficial across these training configurations.
This single-seed comparison does not establish the best weight for every
backbone.

\section{Open-Vocabulary SED Detector Baselines}
\label{app:detectors}

We evaluate off-the-shelf open-vocabulary detectors on the same query and interval schema as the LALMs. Their metric profiles appear in Table~\ref{tab:main_results}; Table~\ref{tab:riou_sweep} adds DASM's results at stricter and looser IoU thresholds.

\paragraph{CLAP.}
We use the archived sliding-window results of \texttt{laion/clap-htsat-unfused}, with five-second windows and one-second hops. The recorded cosine thresholds are $0.2$ on AEGBench and DESED, and $0.5$ on AudioGrounding. Stored intervals are rescored with the same one-to-one interval criterion as the other baselines. The AEGBench main comparison and target-conditioned difficulty breakdown use the released query inventory, with CLAP threshold $0.2$. GO/NS/SA retain recording-context membership.

\paragraph{DASM.}
DASM encodes the query text with an MGA-CLAP encoder into a query embedding and
produces frame-level sigmoid activations with DASM\_HTSAT ($32$\,kHz, $64$ mel
bins, hop $320$, window $1024$). Long clips are processed in $10$\,s windows with
a $5$\,s hop (overlapping frames merged by maximum), followed by a median filter
(size $16$) and a softmax temperature $w{=}0.5$. We run DASM at its default
detection threshold of $0.5$ (minimum duration $0.3$\,s).

At its default threshold, DASM favors precision over recall, limiting event recovery relative to the adapted LALMs. The full difficulty-subset mIoU comparison, including the Qwen3-Omni adaptation stages, appears in Table~\ref{tab:hardcase_full}.

\paragraph{Target-conditioned recovery and interval matching.}
Table~\ref{tab:hardcase_riou} complements the target-conditioned mIoU columns in Table~\ref{tab:hardcase_full}. PE-A-Frame has higher overlap and recovery on repeated targets, whereas CLAP has higher mIoU and R@0.5 on sustained targets. Auto-AEG has the highest micro Interval F1@0.5 on overlap, repetition, and sustained-event queries, showing the distinction between best-match coverage and one-to-one instance matching.

\begin{AppendixTable}
\centering\footnotesize
\setlength{\tabcolsep}{2pt}
\begin{tabularx}{\linewidth}{@{}l*{6}{C}@{}}
\toprule
& \multicolumn{3}{c}{\textbf{R@0.5}} & \multicolumn{3}{c}{\textbf{IF1}} \\
\cmidrule(lr){2-4}\cmidrule(l){5-7}
Model & PO & RO & LD & PO & RO & LD \\
\midrule
Gemini-3-Pro & 0.251 & 0.136 & 0.356 & 0.198 & 0.146 & 0.224 \\
Kimi-Audio-7B & 0.068 & 0.095 & 0.023 & 0.017 & 0.038 & 0.009 \\
Qwen2-Audio-7B & 0.077 & 0.082 & 0.010 & 0.065 & 0.079 & 0.018 \\
Audio Flamingo 3 & 0.153 & 0.169 & 0.082 & 0.044 & 0.073 & 0.008 \\
CLAP & 0.239 & 0.039 & \textbf{0.587} & 0.180 & 0.050 & 0.320 \\
TAG & 0.183 & 0.286 & 0.134 & 0.155 & 0.237 & 0.028 \\
DASM & 0.182 & 0.219 & 0.148 & 0.250 & 0.267 & 0.099 \\
PE-A-Frame & 0.335 & \textbf{0.372} & 0.412 & 0.204 & 0.241 & 0.025 \\
TEMPO SFT & 0.317 & 0.273 & 0.323 & 0.175 & 0.188 & 0.078 \\
TEMPO + GRPO & 0.364 & 0.306 & 0.449 & 0.245 & 0.259 & 0.152 \\
Q3 base & 0.207 & 0.116 & 0.396 & 0.171 & 0.130 & 0.254 \\
Q3 + SFT & 0.288 & 0.105 & 0.013 & 0.185 & 0.110 & 0.023 \\
Q3 + GRPO & \textbf{0.420} & 0.339 & 0.569 & \textbf{0.368} & \textbf{0.366} & \textbf{0.376} \\
\bottomrule
\end{tabularx}
\caption{Target-conditioned recovery (query-macro R@0.5) and micro Interval F1@0.5 on the complete 9,924-query AEGBench evaluation set.}
\label{tab:hardcase_riou}
\end{AppendixTable}

\section{Data Appendix}
\label{app:data}

This appendix consolidates the contents and format of the two
novel resources introduced in the paper: the \textbf{Auto-AEG} training corpus
and the \textbf{AEGBench} evaluation benchmark. Their construction procedures
are detailed in the construction sections (Stage~1 Synthesis, Stage~2
Annotation, and AEGBench Construction); here we summarize their statistics for
reproducibility (Table~\ref{tab:data_stats}).
Only the derived annotation files are planned for distribution; the original
third-party audio is excluded.

\paragraph{Auto-AEG.}
Auto-AEG pairs synthetic clips carrying placement-exact ground truth with real recordings carrying multi-model pseudo-labels. Table~\ref{tab:data_stats} consolidates corpus sizes; Appendices~\ref{app:synth} and~\ref{app:annot} describe the two construction stages.

\paragraph{AEGBench.}
AEGBench provides independently annotated List-All queries from four audio sources, disjoint from the training splits. Its source selection, non-exclusive difficulty tags, and boundary reconciliation are detailed in Appendices~\ref{app:bench} and~\ref{app:agreement}.

\paragraph{Record schema.}
Every record follows a uniform schema. The audio is a single-channel 16\,kHz
waveform; the query is a natural-language List-All prompt naming one
sound category; the gold answer is a JSON array of $[t_{\text{start}},
t_{\text{end}}]$ pairs in seconds (an empty array denotes absence). Auto-AEG
Stage~1 records additionally store the occurrence count and the per-segment mix
parameters used to synthesize the clip; AEGBench records additionally store
their source provenance and the assigned difficulty tags.
Figure~\ref{fig:train_sample} illustrates a Stage~2 record.

\begin{AppendixFigure}
\centering
\begin{minipage}{\columnwidth}
\begin{Verbatim}[fontsize=\footnotesize,breaklines=true,breakanywhere=true]
{
  "audio": "example.flac",
  "duration_sec": 11.941,
  "caption": "Strings, brass, percussion and synth.",
  "categories": ["string section", "brass instrument",
                 "percussion", "synthesizer"],
  "clips": [
    {"category": "string section", "start": 0.0, "end": 11.2},
    {"category": "brass instrument", "start": 0.0, "end": 11.8},
    {"category": "percussion", "start": 0.0, "end": 11.941},
    {"category": "synthesizer", "start": 0.0, "end": 11.941}
  ]
}
\end{Verbatim}
\end{minipage}
\caption{Illustrative Stage~2 record, not an actual data item. Each category defines a List-All query supervised by its intervals. The original caption is retained for reference and is not used as supervision.}
\label{fig:train_sample}
\end{AppendixFigure}

% \paragraph{Release and licensing.}
% Upon acceptance, the Auto-AEG corpus and the AEGBench benchmark (including the
% audio, the query/answer pairs, the difficulty tags, and the data-construction
% and evaluation scripts) will be released under a \textbf{CC-BY-4.0} license,
% with all accompanying source code under an \textbf{MIT} license, via a public
% repository. No personally identifying information is included, and audio sourced
% from third parties is redistributed under the terms of its respective license.
% The release is provided solely to enable free usage for research purposes.

\begin{AppendixTable}
  \centering
  \small
  \begin{tabular}{@{}lrr@{}}
    \toprule
    \textbf{Resource} & \textbf{Clips} & \textbf{Queries} \\
    \midrule
    \multicolumn{3}{l}{Auto-AEG (training corpus)}\\
    \quad Synthetic (SFT)     & 10{,}000 & 10{,}000 \\
    \quad Real (GRPO)         & 1{,}987  & 4{,}910  \\% &  2{,}000 &  5{,}244 \\
    % \quad \textbf{Total}                        & \textbf{12{,}000} & \textbf{15{,}244} \\
    \midrule
    \multicolumn{3}{l}{AEGBench (evaluation benchmark)}\\
    \quad Total               & 3{,}425  & 9{,}924  \\
    % \quad \textbf{Total}                         & \textbf{3{,}425} & \textbf{9{,}924} \\
    \bottomrule
  \end{tabular}
  \caption{Consolidated statistics for the two novel resources. Auto-AEG is the
    training corpus after training-time filtering (placement-exact synthetic labels and real-audio pseudo-labels); AEGBench is the
    difficulty-stratified evaluation benchmark.}
  \label{tab:data_stats}
\end{AppendixTable}

% \section{AI Assistance Usage}
% In this paper, we used AI agents to assist with the literature review, manuscript formatting and polishing, and the execution and scheduling of experimental scripts.

\end{document}